\documentclass[twocolumn,showpacs,prr,aps,longbibliography,superscriptaddress]{revtex4-2}
\usepackage[bookmarks=false,linkcolor=blue,urlcolor=blue,colorlinks,citecolor=blue]{hyperref}
\usepackage{tabularray}
\pdfoutput=1

\usepackage{graphicx} %
\usepackage{amsmath}
\usepackage{xcolor}
\usepackage{placeins}
\definecolor{pink}{rgb}{1,0,0.9}

\usepackage{amssymb}   
\usepackage{algorithm}         

\makeatletter
\makeatletter
\expandafter\edef\csname ftype@algorithm\endcsname{\the\numexpr\value{float@type}/2\relax}
\makeatother

\usepackage{algpseudocode}     
\usepackage{braket}            

\algnewcommand\algorithmicparfor{\textbf{parallel for}}
\algdef{SE}[PARFOR]{ParFor}{EndParFor}[1]%
  {\algorithmicparfor\ #1\ \algorithmicdo}%
  {\algorithmicend\ \algorithmicparfor}

\newcommand{\Nbw}{N_{\mathrm{bw}}}
\newcommand{\Phase}[1]{\Statex\hspace{-\algorithmicindent}\textit{\color{black!65}#1}}
\usepackage{tikz}
\usetikzlibrary{arrows.meta,  
                calc}         

\definecolor{acc}{RGB}{31,119,180}    
\definecolor{acc2}{RGB}{214,96,26}    

\begin{document}
\title{Efficient computation of PEPS gradients for optimized quantum state preparation on 2D topologies.}
\author{Illia Lukin}
\affiliation{Haiqu, Inc., New York City; Newlab, 19 Morris Ave, Brooklyn, NY 11205, USA}
\affiliation{Akhiezer Institute for Theoretical Physics, NSC KIPT, Akademichna 1, 61108 Kharkiv, Ukraine}
\author{Vladyslav Bohun}
\author{Maciej Koch-Janusz}
\affiliation{Haiqu, Inc., New York City; Newlab, 19 Morris Ave, Brooklyn, NY 11205, USA}

\begin{abstract}
Quantum state preparation is a fundamental subroutine for quantum computation. For complex, entangled initial states it is, however, difficult to prepare shallow quantum circuits approximating the state with sufficient accuracy. Tensor networks provide a promising avenue of synthesis of such circuits, but most approaches employ 1D Matrix Products States, which struggle to accurately express entangled states in higher dimensions. Here we present a hardware topology-aware quantum state preparation method based on gradient optimization of 2D projected entangled pair states (PEPS). Our approach combines local disentanglement, with global fine tuning based on PEPS gradient optimization. Towards this end we describe an efficient procedure for PEPS gradient computation. We test the convergence and performance of our method in numerical simulations.
\end{abstract}

\maketitle

\section{Introduction}

Quantum state preparation is not only the first step of most quantum algorithms, but also one whose complexity may invalidate any putative quantum advantage. Applications such as quantum computational fluid dynamics \cite{Jaksch2023}, more general quantum PDE solvers \cite{Childs2021}, quantum chemistry \cite{Fomichev2024}, condensed matter \cite{LinTong2020} or quantum finance algorithms ~\cite{Stamatopoulos2020, Iaconis2024} require or benefit from initalisation with a carefully selected quantum state. However, for an arbitrary state the cost of exact preparation may even scale exponentially in the number of qubits \cite{Shende2006}. Fortunately, many states of interest can be approximated to a high accuracy using families of quantum states, which themselves require preparation circuits of only polynomial depth. An important such class are matrix product states (MPS), which admit a number of different efficient circuit synthesis methods, including those based on renormalization group ~\cite{Malz2024, Scheer2025, Wei2026}, isometric canonical form of MPS ~\cite{Schon2005, Ran2020, Rudolph2023}, disentanglement ~\cite{Mansuroglu2026} and variational optimization ~\cite{Green2025, Szoldra2026}. Quantum state preparation based on MPS was used in a variety of different applications, including function preparation ~\cite{Holmes2020, GarciaRipoll2021, Bohun2026}, distribution loading ~\cite{Iaconis2024} and image loading ~\cite{Jobst2024, Dilip2022}. 


While MPS can accurately represent 1D short-range correlated states, they struggle in higher dimensions \cite{Stoudenmire2012}. Furthermore, they are not well suited to the 2D topology of most quantum processors \cite{Tindall2024} and thus, not only do they fail to accurately represent higher dimensional states, but also incur large gate overheads when transpiled to realistic quantum processing unit (QPU) topologies. 

The most direct generalization of MPS to higher dimensional lattices are the projected entangled pair states (PEPS) ~\cite{Orus2014, Cirac2021}, finding applications in simulations of quantum experiments \cite{Tindall2024, Patra2024, Rudolph2025}, frustrated systems ~\cite{CorbozMila2013, Liao2017}, high-temperature superconductivity ~\cite{Corboz2011, Corboz2014} and in lattice gauge theories~\cite{Emonts2020}. More importantly, PEPS can be applied to state preparation. While recent work~\cite{Szoldra2026} used local disentanglement to synthesize quantum circuits from a PEPS, following similar MPS algorithms~\cite{Mansuroglu2026}, the resulting circuit is generally not globally optimal at a given number of gates and circuit geometry.

Here we develop an improved PEPS-based quantum state preparation algorithm, combining the disentanglement approach with a variational gradient-based optimization of the obtained circuit. A key ingredient is an efficient numerical method for PEPS gradient computation using tensor network contractions, which we introduce. The gradients thus obtained allow to globally optimize (for a given circuit layout) the state preparation circuit, which can be used in \emph{e.g.}~variational quantum eigensolver initialization on 2D lattices, or to prepare initial states in 2D quantum quench simulations, among others. We validate the properties of the improved circuits and the convergence of the method in numerical simulations on the examples of the 2D Heisenberg and kicked Ising models.

The rest of the manuscript is structured as follows: in Sec.~\ref{sec:main_alg} we briefly review PEPS as well as the disentanglement method of circuit preparation. In Sec.~\ref{sec:gradients} we describe our gradient-based PEPS based state preparation, and particularly the procedure for efficient PEPS gradient computation. In Sec.~\ref{sec:numQPU} we numerically study the performance and convergence of the method. We conclude in Sec.~\ref{sec:conclusions}. The appendix gives further technical details.

\section{PEPS-distentanglement and quantum state preparation}
\label{sec:main_alg}



\begin{figure*}
\centering
\includegraphics[width= 0.85\linewidth]{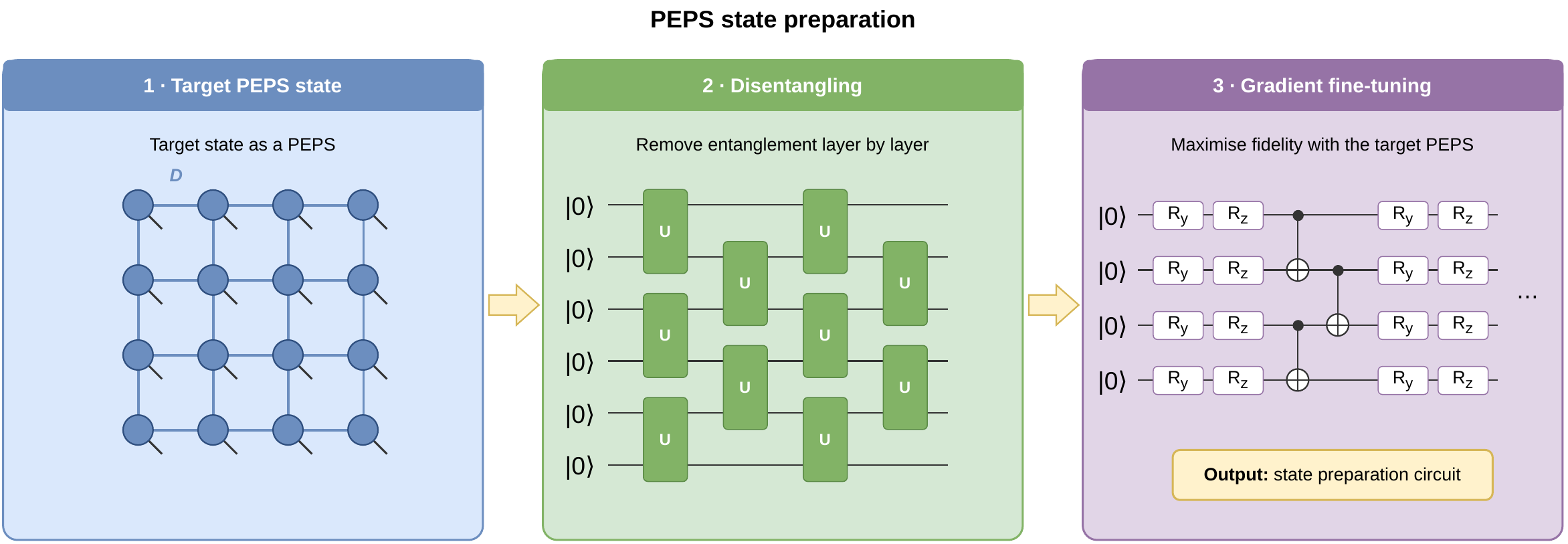} 
 \caption{\label{fig:peps_preparation}%
   An overview of our PEPS-based quantum state state preparation. An input PEPS state, the result of a classical simulation, is synthesized as a shallow quantum circuit in two steps. First, a local disentanglement procedure builds the circuit following a pre-determined ansatz topology layer by layer, in a greedy fashion (see Sec.~\ref{sec:main_alg}). In the second step all of the gates' parameters are trained with a global approach using an efficient PEPS gradient algorithm we describe in Sec.~\ref{sec:gradients}.}
\end{figure*}

Here we describe the disentanglement approach to constructing quantum circuits, which in Sec.~\ref{sec:gradients} we improve with gradient fine-tuning (see Fig.~\ref{fig:peps_preparation}). Our procedure for this first stage, the PEPS disentanglement, is similar to that of Refs.~\cite{Mansuroglu2026, Szoldra2026}. The input is assumed to be a PEPS state, obtained, \emph{e.g.}, as a result of classical PEPS-based simulation of the ground state of a quantum Hamiltonian.


\subsection{PEPS}

\begin{figure*}
\includegraphics[width= 0.75\linewidth]{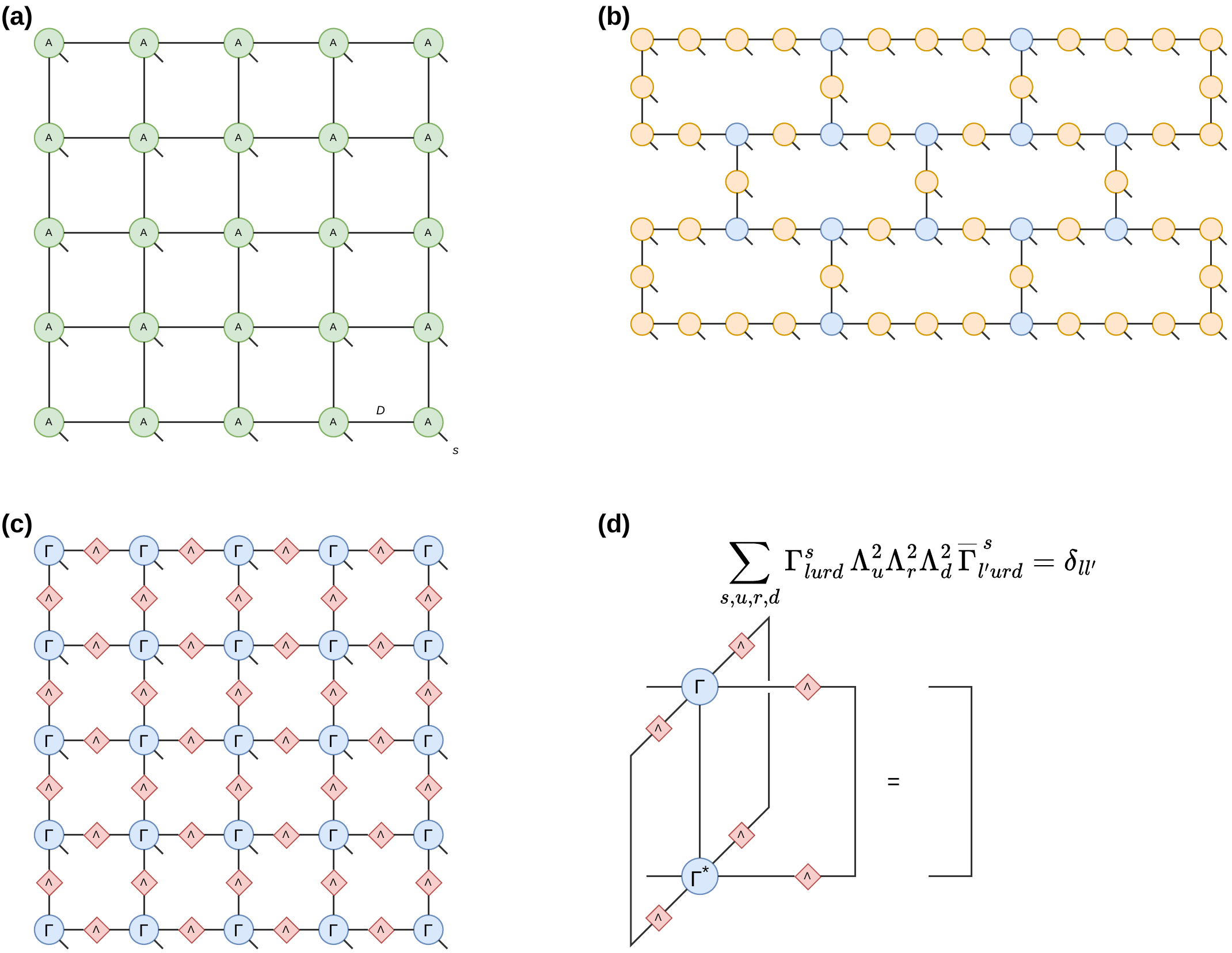} 
 \caption{\label{fig:peps_definition}%
   Different PEPS topologies: (a) PEPS on a square lattice. (b) PEPS on a heavy-hex lattice. (c) PEPS on a square lattice in the superorthogonal canonical form. As is standard, open legs represent physical indices, and the interior edges represent contracted auxiliary indices of dimension $D$ of tensors at either sites or edges of the lattice (d) Additional relations between the $\Gamma$ tensors and bond matrices $\Lambda$ in the superorthogonal canonical form. Analogous relations hold for the other legs.}
\end{figure*}

Projected entangled pair states ~\cite{Orus2014, Cirac2021} are a natural generalization of matrix product states ~\cite{Schollwock2011} to 2D and 3D lattices, although here we focus on the 2D case. For a given planar graph whose vertices are labeled by $i$, \emph{e.g.}~a square lattice, we place a local Hilbert space on each vertex, whose basis states are labeled by the physical index $s_i$. In our case these correspond to $\ket{0}$ and $\ket{1}$ states of the qubit. The PEPS wave function is then constructed as follows: at every graph vertex $i$ of valency $z$ there is a rank $(z+1)$ tensor $A^{s_i}_{\alpha_{1} \alpha_{2} ... \alpha_{z}}$. The auxiliary or virtual bond variables $\alpha$ are associated with the edges incident at the vertex and take values in ${{1, 2,..., D}}$, where $D$ is the PEPS bond dimension. The amplitude of a quantum state corresponding to a physical configuration $\ket{s_1,s_2,\ldots}$ is given by contracting the pairs of virtual indices of tensors $A^{s_i}_{\alpha_{1} \alpha_{2} ... \alpha_{z}}$ corresponding edges shared between pairs of vertices. In Figs.~\ref{fig:peps_definition}(a-b) we depict PEPS wavefunctions for square and heavy-hex lattices, where the site tensor are represented by graph nodes, the contracted virtual indices are edges linking the vertices and the physical degrees of freedom are depicted by uncontracted open lines.

PEPS naturally approximate area law entanglement states~\cite{Eisert2010}: for a given bond dimension $D$ the entanglement entropy of a spatial region with boundary length $L$ is bounded from above by $L \times \log{D}$. Many important classes of states, particularly ground states of gapped local Hamiltonians fall in this category. For some of them, such as the ground states of the toric code, RVB liquid or 2D AKLT model exact PEPS representations are known~\cite{Verstraete2004, Verstraete2006, Schuch2010, Schuch2012}. Here, we exploit the capability of PEPS to accurately approximate the results of application quantum circuits of moderate depths to simple initial states. Most of quantum states which are possible to prepare on NISQ quantum devices are of such nature~\cite{Tindall2024, Patra2024}.



In what follows we use the superorthogonal canonical form of PEPS ~\cite{Ran2012, Phien2015recycling}, to which any PEPS can be transformed~\cite{Tindall2023}. In this form the node tensors are factorized in such a way that we now have node tensors $\Gamma^{s_i}_{\alpha_{1} \alpha_{2} ... \alpha_{z}}$ and, additionally, on each edge a diagonal positive matrix $\Lambda_{\alpha} \delta_{\alpha \beta}$, where $\alpha, \beta$ are the virtual indices of two nodes adjacent to the bond and $\delta$ is the Kronecker symbol. This is depicted in Fig.~\ref{fig:peps_definition}(c). The PEPS wavefunction is, as before, obtained by contracting all virtual indices of site and bond tensors, leaving only the physical indices. Superorthogonality implies that the site and bond tensors obey certain isometric conditions, which are graphically represented in Fig.~\ref{fig:peps_definition}(d).

The entries of the diagonal bond matrices $\Lambda_{\alpha}$ are useful in constructing truncated approximations, as they are related to entanglement. When the underlying graph is a tree, they are explicitly the Schmidt coefficients of the bipartition of the quantum state obtained by cutting the corresponding bond. On graphs with loops, this exact correspondence no longer holds true, but if the correlations are sufficiently tree-like it is possible to use truncation of the entries of $\Lambda_{\alpha}$ as an approximation. This approach is the basis of Simple Update~\cite{Jiang2008} based algorithms and Belief Propagation \cite{Tindall2023} based algorithms,  which we also employ.

\subsection{Disentanglement}

Given a PEPS state our goal is to find a quantum circuit preparing it by applying a quantum circuit to an initial product state $|00...0\rangle$. The first stage of constructing this circuit is a local disentanglement procedure~\cite{Mansuroglu2026, Szoldra2026} which is greedy, and whose suboptimal result we will improve upon in Sec.~\ref{sec:gradients}.


To make the procedure concrete, we first need to fix the architecture of the resulting quantum circuit. Targeting devices with limited connectivity, such as superconducting QPUs, we allow only nearest-neighbour gates in the QPU topology, and assume a dense brick-wall arrangement of the disentangling gate blocks on the edges of this graph, or its subgraph. Each such block is a parametrised subspace of $SU(4)$-unitaries acting on the two qubits sharing an edge, and there are multiple layers thereof. We employ a parametrised circuit ansatz shown in Fig.~\ref{fig:brick-wall}(b). The disentangling procedure described below fixes the circuit parameters. As it is entirely local, blocks in a single disentangling layer can be optimised in parallel, as long as they act on disjoint qubits. Maximal sets of such blocks can be found by solving the bond colouring problem (see Fig.~\ref{fig:brick-wall}(a)). Finally, we add a single layer of 1-qubit gates, which we will discuss later.

\begin{figure}
\includegraphics[width= 0.8\linewidth]{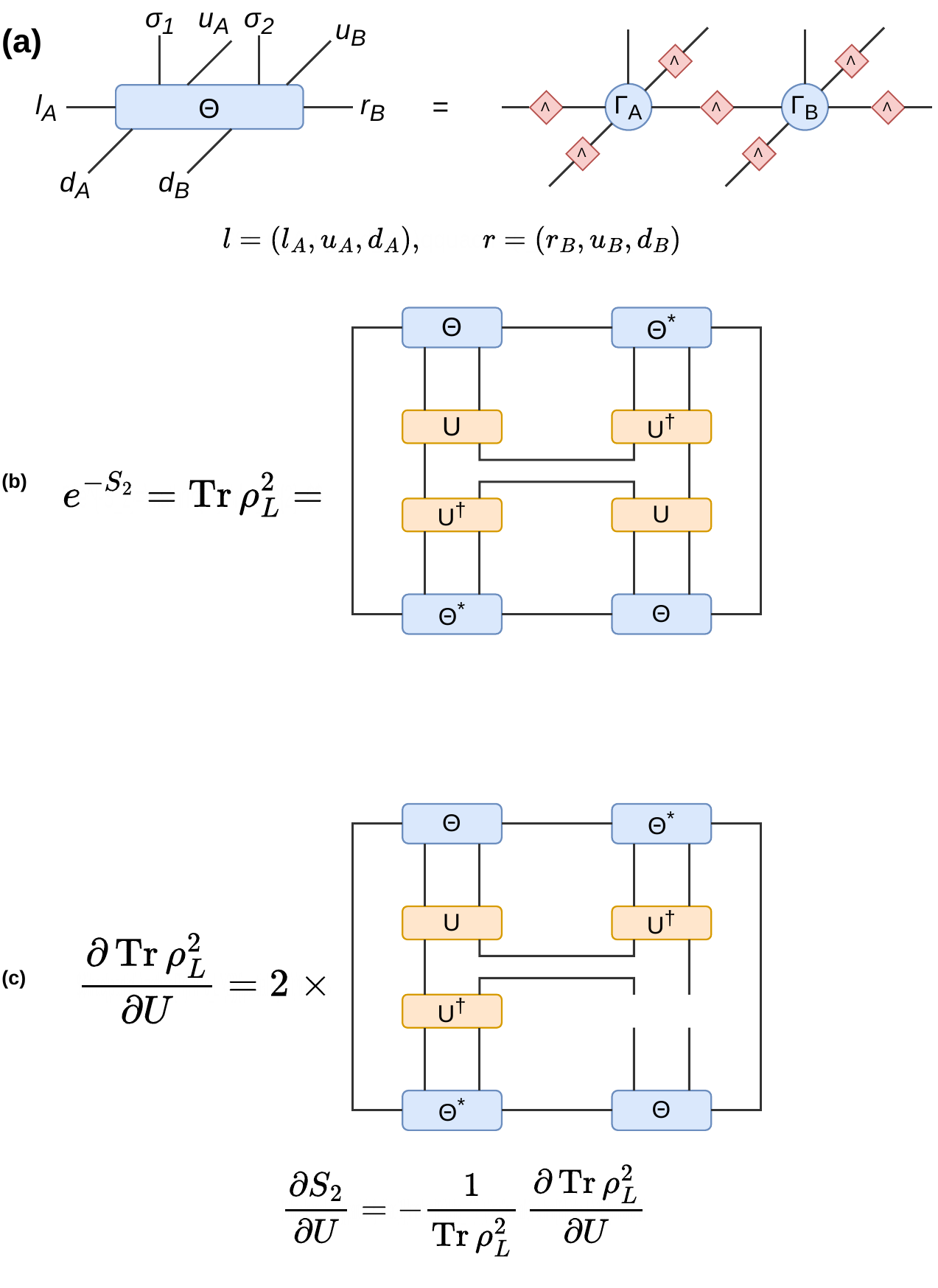} 
 \caption{\label{fig:disentangling}%
   Disentangling a bond with unitary matrix $U$ applied to the physical indices at neighbouring sites: (a) We form a Simple Update approximation $\Theta$ for the purification of the local reduced density matrix on the bond. (b) From $\Theta$ we form the approximate Rényi entropy $S_{2}$ for the bond, computed after the disentangling gate $U$ is applied. (c) The gradient of the entropy $S_{2}$ with respect to $U$, used in $U$ optimization.}
\end{figure}

The parameters of the quantum circuits are determined using the following local disentangling approach~\cite{Hauschild2018, Mansuroglu2026, Szoldra2026}. For each layer, consider the set of parametrised blocks $U$, which can be optimised in parallel (a sublayer of single colour class of Fig.~\ref{fig:brick-wall}(a)) and which are applied to the input PEPS state. For each of them we formulate an independent, local variational problem on the corresponding lattice edge, which is to find parameters of $U$ minimizing the bond 2-R\'enyi entropy $S_{2} = -\ln{\sum_{\alpha} \Lambda_{\alpha}^{4}}$ (with normalization $\sum_{\alpha} \Lambda_{\alpha}^{2} = 1$) on that bond. In essence, we are asking that each parametrised gate block $U$ locally remove as much entanglement as possible from the input state. This local cost function can be computed by performing the tensor contraction shown in Fig.~\ref{fig:disentangling}(a-b). The gradients of this loss can also be easily computed, as shown in Fig.~\ref{fig:disentangling}(c). In practice we use them with the L-BFGS optimizer \cite{Nocedal1980, Liu1989}. After finding the gate parameters we apply the gate to the input using the Simple Update algorithm~\cite{Jiang2008}, and update the PEPS state. We repeat the procedure for the next colour sublayer, and the following layers.


The above cost function is motivated by the fact that when the PEPS graph is a tree, the bond R\'enyi entropy is the R\'enyi entropy of entanglement bipartition, and minimizing it is optimal in the sense of removing entanglement from the state. At the end of such procedure the input PEPS state is brought to a simple product state by application of the optimised gate blocks $U$, and thus the conjugate of the disentangling circuit creates the input PEPS state. Moreover, the application of gate blocks to the PEPS does not change the $\Lambda$ matrices on other bonds, and so the optimization of gates on different bonds in a layer can be done in parallel. Finally, when using Simple Update~\cite{Jiang2008} to apply gates, very small diagonal elements of $\Lambda$ obtained after accurate disentanglement can be safely truncated. When all entries but the leading ones are below a very small truncation threshold, the state is close to a product state, which can be transformed to $|00...0\rangle$ with application of one layer of $1$-qubit gates. 


This last 1-qubit gate layer is obtained using local $1$-qubit reduced density matrices, which can either be computed approximately with Simple Update environments or in a controlled accurate way using either boundary MPS (bMPS) contractions or loop expansions~\cite{Evenbly2026}. The easily computed leading eigenvector of such reduced density matrix corresponds to the most probable state obtained by measuring the corresponding qubit, and the 1-qubit unitary is found as the one mapping it to the zero state $|0\rangle$.


\begin{algorithm}[t]
\caption{Greedy disentanglement of a PEPS}\label{alg:dis}
\begin{algorithmic}[1]
\small\raggedright
\Require Target PEPS state $\ket{PEPS}$ in a superorthogonal form $\{\Gamma,\Lambda\}$ (Fig.~\ref{fig:peps_definition}(c,d)); bond coloring $\mathcal E_1,\dots,\mathcal E_c$ (Fig.~\ref{fig:brick-wall}(a)); number of brick-wall layers $\Nbw$
\Ensure disentangler $U(\theta_0)$ with $U(\theta_0)\ket{PEPS}\approx\ket{0\cdots0}$
\For{$p=1,\dots,\Nbw$}\Comment{brick-wall layers}
  \For{$i=1,\dots,c$}\Comment{sublayers}\label{ln:dis-sub}
    \ParFor{bond $b=(A,B)\in\mathcal E_i$}
      \State $\Theta_b\gets$ two-site purification on $b$\Comment{Fig.~\ref{fig:disentangling}(a)}\label{ln:dis-theta}
      \State $U_b\gets\arg\min_U S_2(U\Theta_b)$\Comment{Fig.~\ref{fig:disentangling}(b,c)}\label{ln:dis-opt}
    \EndParFor
    \State apply $\{U_b\}_{b\in\mathcal E_i}$ to $\ket{PEPS}$; \label{ln:dis-apply}
  \EndFor
\EndFor
\ForAll{sites $j$}\Comment{final 1-qubit layer}\label{ln:dis-final}
  \State $\rho_j\gets$ 1-qubit reduced density matrix\label{ln:dis-rdm}
  \State $v_j\gets$ eigenvector of $\rho_j$ with largest eigenvalue
  \State $W_j\gets R_Y R_Z$ such that $W_j v_j\propto\ket 0$\label{ln:dis-w}
\EndFor
\State \Return $U(\theta_0)=\big(\textstyle\bigotimes_j W_j\big)\,U^{(\Nbw)}\cdots U^{(1)}$, with\par
   \hskip\algorithmicindent $U^{(p)}=\big(\prod_{b\in\mathcal E_c}U_b\big)\cdots\big(\prod_{b\in\mathcal E_1}U_b\big)$
\end{algorithmic}
\end{algorithm}

The full local disentanglement procedure is summarised in Algorithm~\ref{alg:dis}. Each brick-wall layer is split into sublayers of non-intersecting bonds (line~\ref{ln:dis-sub}). The gates of one sublayer are optimised independently, by minimising the local bond R\'enyi entropy with L-BFGS (line~\ref{ln:dis-opt}), and then applied to the PEPS with Simple Update or Belief Propagation~\cite{Jiang2008, Patra2024, Tindall2023}, or optionally with a more accurate method such as Full Update~\cite{Czarnik2018, Czarnik2019, Dziarmaga2021, Dziarmaga2022, Hubig2019, Ponnaganti2023} (line~\ref{ln:dis-apply}). After all brick-wall layers are processed, a final layer of 1-qubit gates maps the state to $\ket{0\cdots0}$ (lines~\ref{ln:dis-final}--\ref{ln:dis-w}). To construct it, we compute the 1-qubit reduced density matrices, either approximately from Simple Update environments or in a controlled way with bMPS contractions or loop expansions~\cite{Evenbly2026}, and rotate the eigenvector with the largest eigenvalue, i.e.\ the most probable measurement outcome on that qubit, to $\ket 0$ with a product of $R_Y$ and $R_Z$ gates.


\begin{figure*}[!th]
\includegraphics[width= 0.8\linewidth]{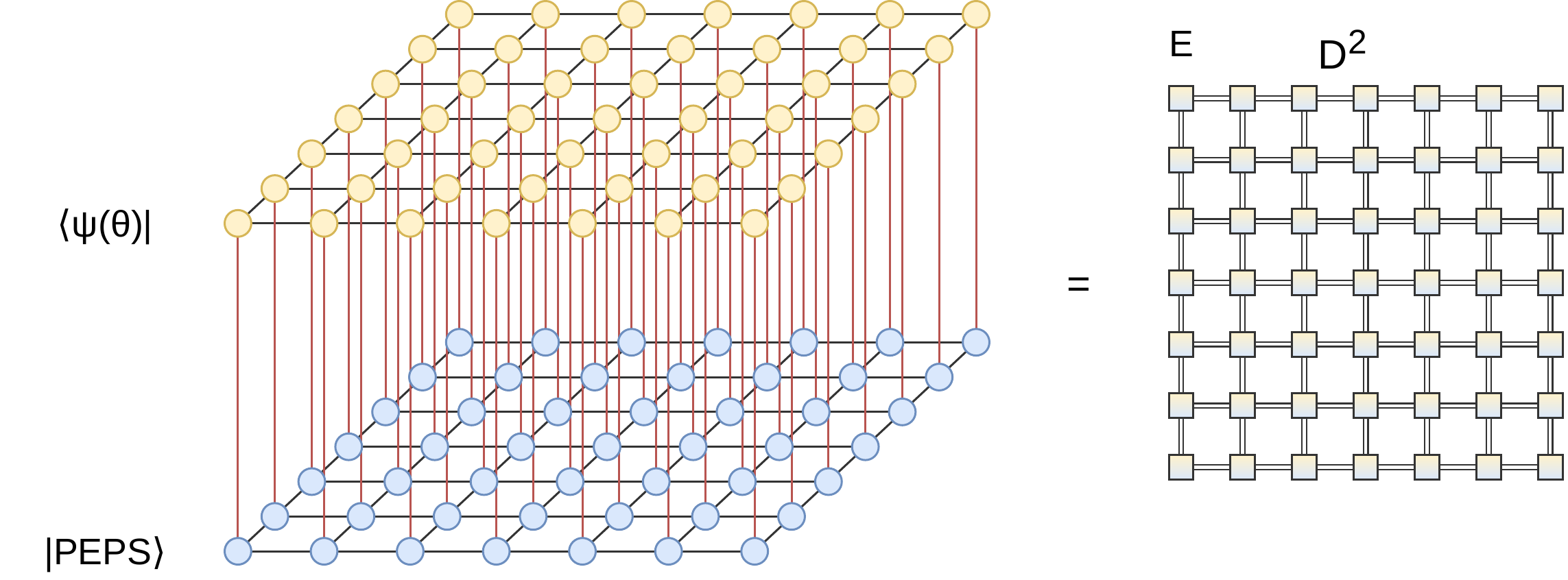} 
 \caption{\label{fig:peps_overlap}%
   The overlap between two PEPS states: the input $|PEPS\rangle$ and a PEPS representation of the  parametrised disentagler approximation thereof $\langle \psi(\theta)|$. The overlap tensor network is built from two-layer tensors with an effective bond dimension $D^{2}$.  }
\end{figure*}

\section{Circuit optimisation with PEPS gradients} 
\label{sec:gradients}
\subsection{Fidelity optimization}

The above local disentanglement procedure is greedy, and for a given circuit architecture cannot produce globally optimal parameter values. Consequently, the obtained quantum circuit is not the most accurate approximation of the input state (see Sec.~\ref{sec:numQPU}). To produce one as accurate as possible at a given circuit depth -- the central goal of the whole procedure -- we introduce a global approach fine-tuning the parameters of the initial disentangled circuit, which, crucially, is computationally tractable at scale.

Towards this goal we introduce the global infidelity cost
\begin{equation}\label{eq:fidelity}
    L(\theta) = 1 - |\langle 0| U(\theta)|PEPS\rangle|^{2}, 
\end{equation}
where $|PEPS\rangle$ is the target PEPS state, $U(\theta)$ is the disentangling circuit, and $\theta$ described the whole collection of gate angles, initially fixed to the values obtained via local disentanglement. 
In our procedure, we optimise these angles directly and globally, computing gradients of the fidelity with respect to them. This is made possible by two observations. First, even though this is a global cost function, which in principle can exhibit barren plateaus \cite{Cerezo2021}, local disentanglement effectively serves as a warm-start procedure, ensuring the gradients are large initially. Second, we describe a computationally efficient way of approximately evaluating both the cost function and its gradients.

Let us first consider evaluating the cost function. Note that the disentangling circuit can be arbitrarily factorized into $U(\theta) = U_{2}(\theta)U_{1}(\theta)$, and the computation in Eq.~\ref{eq:fidelity} reduces to the scalar product between the states $U_{2}^{\dagger}(\theta)|0\rangle$ and $U_{1}(\theta)|PEPS\rangle$. The extreme cases correspond to first creating a PEPS for the approximate state $U^{\dagger}(\theta)|0\rangle$ and contracting it with the target state $|PEPS\rangle$, on the one hand, or first contracting the target state with the disentangler via $U(\theta)|PEPS\rangle$ and then contracting it with the $|0\rangle$ state.
While mathematically equivalent, these options have different numerical properties. Here, we chose the first method, as shown in Fig.~\ref{fig:peps_overlap}. Based on our experiments, it requires a smaller PEPS bond dimension $D$ (though a larger auxiliary MPS dimension $\chi$, introduced below) to converge.

\begin{figure*}[t]
\includegraphics[width= 0.85\linewidth]{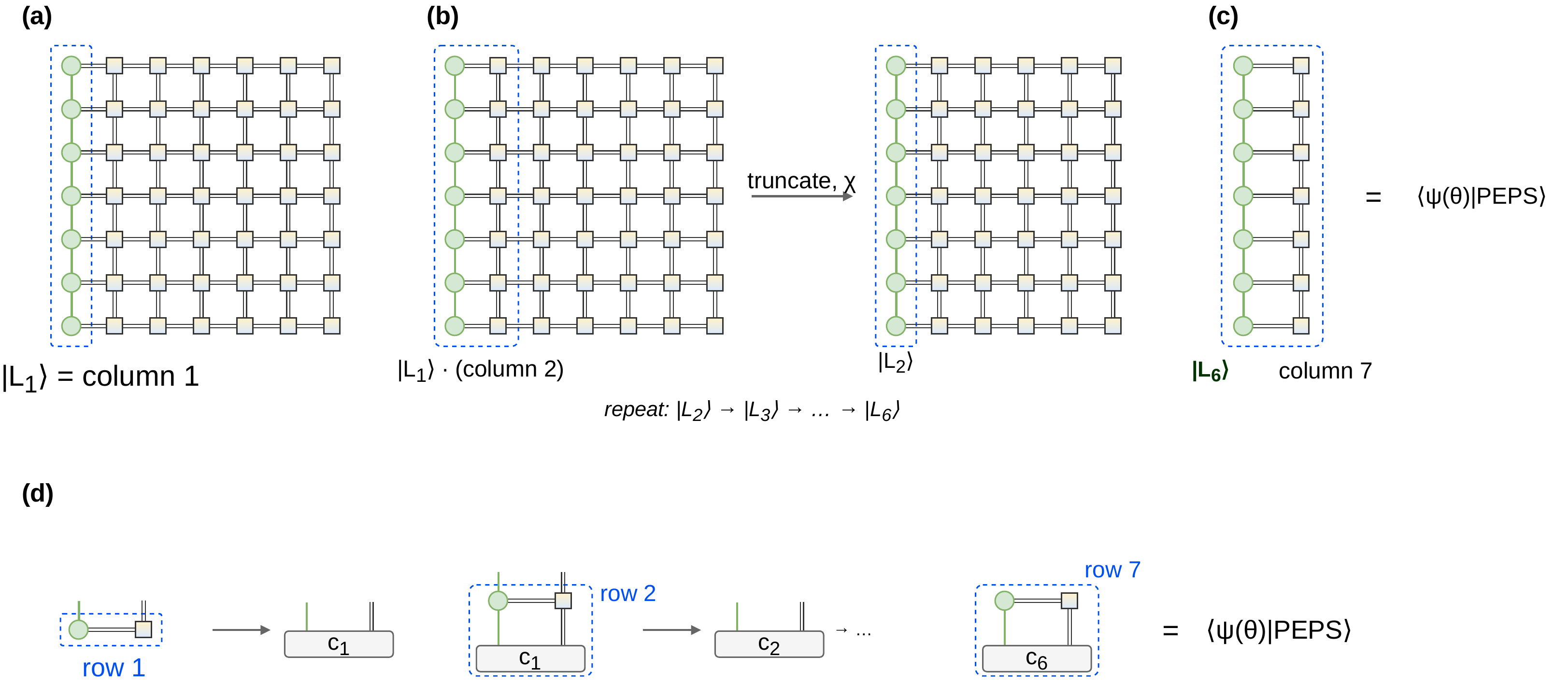} 
 \caption{\label{fig:peps_overlap_bmps}%
   Computing the overlap between the PEPS states $|PEPS\rangle$ and $\langle \psi(\theta)|$ using the bMPS contraction on the example of a $7\times7$ lattice.  (a) Definition of the leftmost bMPS $|L_{1}\rangle$. (b) Contracting sequentially the tensor network columns, treated as MPOs, into the bMPS. (c) The overlap is equal to the product of the final bMPS with the tensors in the last (rightmost) column. (d) The tensor network in (c) is contracted from below, row-wise, using auxiliary row tensors $C_{i}$. }
\end{figure*}

\begin{figure*}[t]
\centering
\includegraphics[width= 1.0\linewidth]{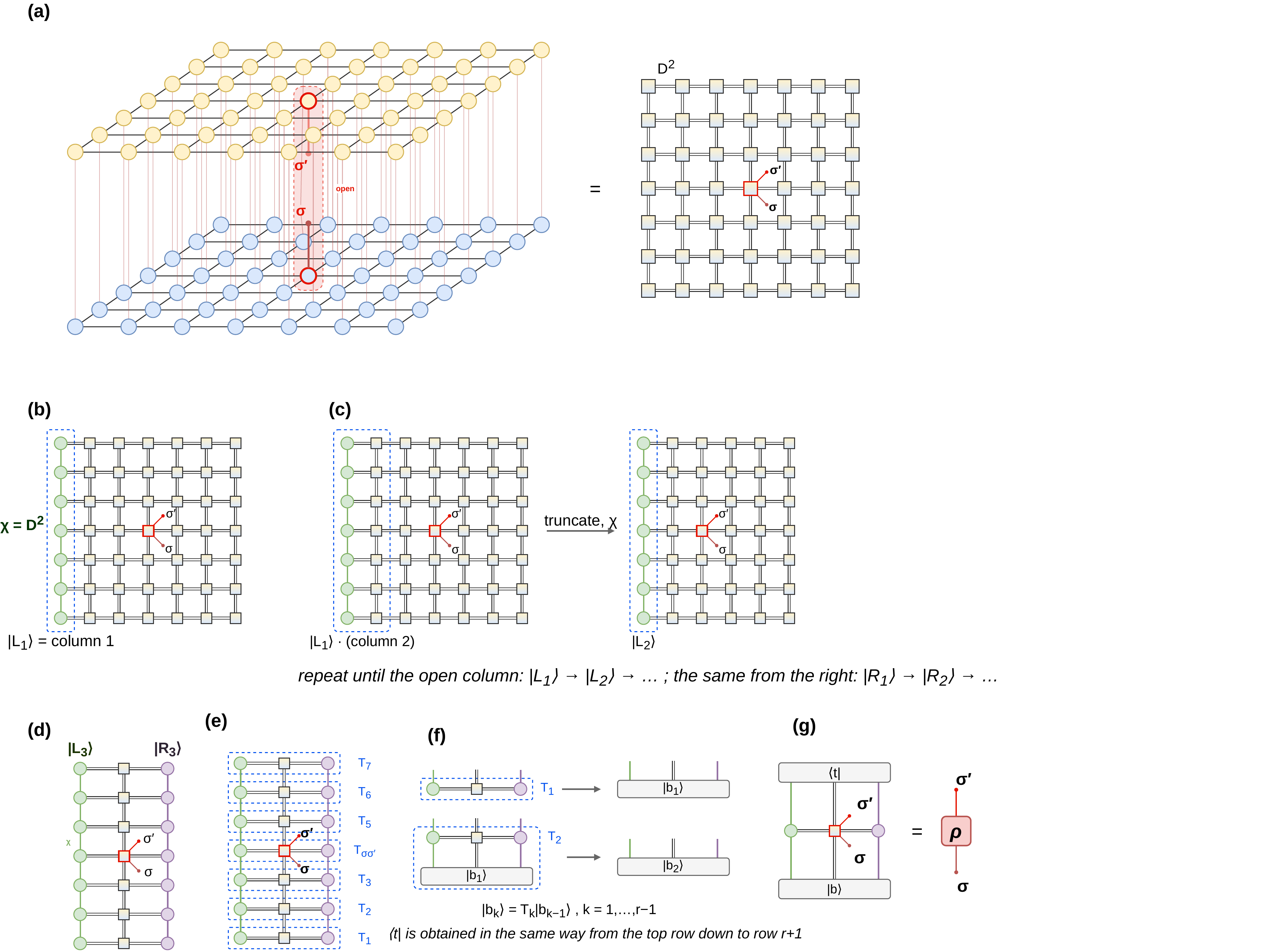} 
 \caption{\label{fig:bmps_contraction}%
   Computing the overlap between different bra and ket PEPS states (with one physical index left open) using a bMPS contraction. (a) Tensor network for the overlap with one open physical index. (b) The leftmost column of the tensor network defines the first left bMPS $|L_{1}\rangle$. (c) The other columns of the tensor network are absorbed into the bMPS, resulting in left bMPS $|L_{k}\rangle$ and right bMPS $|R_{k}\rangle$. (d) The central column with open physical indices is sandwiched between the left and right bMPS. (e) The sandwich tensor network is divided into rows $T_{1}, T_{2}, ...$. (f) The first row $T_{1}$ defines the first bottom tensor $|b_{1}\rangle$. The other bottom tensors $|b_{k}\rangle$ are obtained by contracting the rows: $|b_{k}\rangle = T_{k}|b_{k-1}\rangle$. The top tensors $|t_{k}\rangle$ are defined analogously. (g) The result of the final contraction defines the tensor $\rho$, which is a matrix with two open physical indices.}
\end{figure*}

It is computationally infeasible to exactly contract the PEPS tensors shown in Fig.~\ref{fig:peps_overlap}, as the cost scales as $D^{2L}$, where $L$ is the graph diameter. Fortunately, approximate methods of tensor contraction exist: tensor renormalization group approaches~\cite{Levin2007, Xie2012, Evenbly2015}, methods based on auxiliary tree tensor networks~\cite{Gray2021, Gray2024}, and those employing an auxiliary matrix products states such as bMPS~\cite{Jordan2008, Lubasch2014, Rudolph2025} and CTMRG~\cite{Nishino1996, Orus2009}. Here, we apply the bMPS method, which we adapt to our case and describe in detail.

First, we map the PEPS tensor network on the arbitrary planar graph (corresponding, in our case, to a subgraph of the heavy-hex or square lattice of the QPU) to a square lattice. This is always possible with the inclusion of new vertices or edges, possibly trivial (bond dimension $D=1$)~\cite{Corboz2012}.
The resulting square lattice tensor network can then be contracted using the bMPS approach as shown in Fig.~\ref{fig:peps_overlap_bmps}(a-d): we first divide the square lattice into columns. The leftmost and rightmost columns can be considered a MPS [see Fig.~\ref{fig:peps_overlap_bmps}(a)] with an auxiliary bMPS bond dimension $\chi$, while the middle columns are essentially matrix product operators (MPOs). Starting with the leftmost column $\ket{L_1}$ we apply these MPOs, one by one, to the MPS, truncating the bond dimension back to the fixed value $\chi$ after every step obtaining MPS states $\ket{L_i}$. As the accuracy typically converges exponentially in $\chi$, a moderate $\chi \approx D^{2}$ usually ensures excellent accuracy~\cite{Lubasch2014, Lubasch2014PRB}. We used the canonical form based algorithm~\cite{Schollwock2011, Paeckel2019} for the MPS truncation, and absorbed the bra and ket columns of the tensor networks separately into the boundary MPS, truncating after these separate absorptions. Finally, after all the middle columns are contracted, the final boundary MPS is contracted to the rightmost boundary column MPS $\ket{R}$ [see Fig.~\ref{fig:peps_overlap_bmps}(c)]. This overlap of two MPS can be contracted exactly with the help of auxiliary tensors $C_i$, this time contracting row-wise, as shown in Fig.~\ref{fig:peps_overlap_bmps}(d).
The computational cost of these contractions scales as $O(\chi^{2} D^{5}) + O(\chi^{3} D^{5})$ on the square lattice and $O(\chi^{2} D^{4}) + O(\chi^{3} D^{4})$ on the heavy-hex lattice.


\subsection{Computing PEPS gradients} 

Having discussed loss evaluation, we now describe how to efficiently compute its gradients, which are used in the global optimization of the state preparation circuit. Let $\phi$ be the angle of a $1$-qubit parametrised gate $V(\phi)$. To compute the gradient with respect to it, we factorize the whole disentangler as $U(\theta) = U_{2}(\theta_{2})V(\phi)U_{1}(\theta_{1})$, where the circuit $U_{1}$ consists of the gates which act on $|PEPS\rangle$ prior to $V(\phi)$ and $U_{2}$ consists of those acting after $V(\phi)$. The loss gradient is given by:

\begin{equation}
    \frac{\partial L}{\partial\phi} = - 2 Re \left[ \langle 0| U_{2} \frac{\partial V(\phi)}{\partial \phi} U_{1}|PEPS\rangle \langle 0| U|PEPS\rangle^{*}\right]
\end{equation}

For 1-qubit gates the partial derivative  is given by $\frac{\partial V(\phi)}{\partial \phi} = -iGV(\phi)$, where $G$ is a Hermitian operator, which may also be $\phi$-dependent. For any product of such gates on a single qubit line appearing in the ansatz, \emph{e.g.}~RY and RZ, a parametric expression for $G$ is easily precomputed. The gradient is then rewritten as:

\begin{equation}\label{Eq:gradient}
\begin{split}
    \frac{\partial L}{\partial\phi} = -  2 \left[|\langle 0| U(\theta)|PEPS\rangle|^{2} \right] \times \\ \times Re \left[-i \frac{\langle 0| U_{2} G V(\phi) U_{1}|PEPS\rangle}{ \langle 0| U|PEPS\rangle }\right].
\end{split}
\end{equation}
The first factor is the fidelity, which we derived in the previous section. The evaluation of the second one proceeds in two steps. First, we compute two PEPS tensors corresponding to $|lPEPS\rangle =V(\phi) U_{1}|PEPS\rangle$ and $|rPEPS\rangle =U_{2}^{\dagger}|0\rangle$ (left and right PEPS states). Then, in the second step, 
we contract the physical indices of these two tensors, similarly to computing their scalar product $\langle rPEPS| lPEPS\rangle$, but with one crucial difference: we leave the index on the qubit where $V(\phi)$ acts uncontracted. This is shown in Fig.~\ref{fig:bmps_contraction}(a). The resulting tensor with two physical indices is a $2\times2$ matrix, which we denote by $\rho$, in analogy to the reduced density matrices.

Using the $\rho$ tensor the second factor of the gradient can now be rewritten as $Re \left[ -i \frac{Tr G \rho}{Tr \rho}\right]$. This expression is similar to calculating an observable expectation value with a density matrix, though our $\rho$ is not Hermitian. To compute $\rho$, we again use the bMPS approach: we first perform the column MPS/MPO contractions from the left and from the right, generating a sequence of $\ket{L_i}$ and $\ket{R_i}$ states. We continue until we remain with a tensor expression of the form $\langle L | C | R\rangle$, where $C$ is the column containing the two free physical indices. This, in turn we divide row-wise into tensor networks $T_i$ and contract it with auxuliary tensor states $\ket{t_{j}}, \ket{b_{j}}$ as is shown in detail using standard tensor notation in Fig.~\ref{fig:bmps_contraction}(b-g). Note, that there is no need to recontract the whole tensor network for every gate $V(\phi)$: we first factorize the brick-wall circuit into layers of $1$- and $2$-qubit gates. Then for any layer of $1$-qubit gates all of the gradients are obtained using the same pre-computed boundary MPS, with insertions of the $G$ matrices on different qubits.

\begin{algorithm*}[t]
\caption{Infidelity and all its gradients by tensor-network contraction (Fig.~\ref{fig:sweep}). Costs: square / heavy-hex.}\label{alg:grad}
\begin{algorithmic}[1]
\small\raggedright
\Require target $\ket{PEPS}$; circuit $U(\theta)=U_n\cdots U_1$ over $\{R_Y,R_Z,\mathrm{CX}\}$, grouped into sublayers $U_k$ of non-overlapping gates; $\mathcal K=\{k: U_k \text{ contains parametrised gates}\}$; PEPS and bMPS bond dimensions $D,\chi$
\Ensure loss $\mathcal L(\theta)=1-F$ and $\partial\mathcal L/\partial\phi$ for every angle $\phi$

\Phase{Forward and backward sweeps}
\State $\ket{l_0}\gets\ket{PEPS}$\label{ln:fw0}
\For{$k=1,\dots,n$}\;$\ket{l_k}\gets U_k\ket{l_{k-1}}$, truncate to $D$\Comment{SU/BP, $O(D^{z+1})$ per gate; keep $\ket{l_k}$ for $k\in\mathcal K$}\label{ln:fw}
\EndFor
\State $\bra{r_n}\gets\bra{0\cdots0}$\Comment{product state, $D=1$}\label{ln:bw0}
\For{$k=n,\dots,1$}\;$\bra{r_{k-1}}\gets\bra{r_k}U_k$, truncate to $D$\Comment{keep $\bra{r_k}$ for $k\in\mathcal K$}\label{ln:bw}
\EndFor
\State $F\gets|\langle r_0|PEPS\rangle|^2$,\; $\mathcal L\gets1-F$\Comment{bMPS, Fig.~\ref{fig:peps_overlap_bmps}; more accurate than $|\langle l_n|0\rangle|^2$}\label{ln:loss}
\Phase{Gradients: one ``overlap'' per parametrised sublayer}
\ParFor{$k\in\mathcal K$}\Comment{CX-only sublayers need no contraction}\label{ln:tap}
  \State bMPS $\ket{L_m},\ket{R_m}$ (columns left/right of $m$) of $\langle r_k|l_k\rangle$, all $m$\Comment{Fig.~\ref{fig:bmps_contraction}(b,c); $O(\chi^3D^5)$ / $O(\chi^3D^4)$ per column}\label{ln:bmps}
  \For{each column $m$}
    \State row environments $\ket{b_{j,m}},\bra{t_{j,m}}$ of $\bra{L_m}C_m\ket{R_m}$\Comment{Fig.~\ref{fig:bmps_contraction}(d-f) $O(\chi^3D^4{+}\chi^2D^6)$ / $O(\chi^3D^4{+}\chi^2D^5)$}\label{ln:env}
    \For{each qubit $j$ in column $m$}
      \State $\rho_{jm}\gets\bra{t_{j,m}}\,T^{\sigma\sigma'}_{jm}\,\ket{b_{j,m}}$\Comment{Fig.~\ref{fig:bmps_contraction}(g) $2\times2$, non-Hermitian, $\mathrm{Tr}\,\rho_{jm}=\langle r_k|l_k\rangle$}\label{ln:rho}
      \ForAll{angles $\phi$ of the 1-qubit gates $V_{jm,k}$ on qubit $j$}
        \State $\partial\mathcal L/\partial\phi\gets-2F\,\mathrm{Re}\big[-i\,\mathrm{Tr}(G_\phi\,\rho_{jm})/\mathrm{Tr}\,\rho_{jm}\big]$\Comment{Eq.~(3), with $\partial_\phi V=-iG_\phi V$}\label{ln:grad}
      \EndFor
    \EndFor
  \EndFor
\EndParFor
\end{algorithmic}
\end{algorithm*}

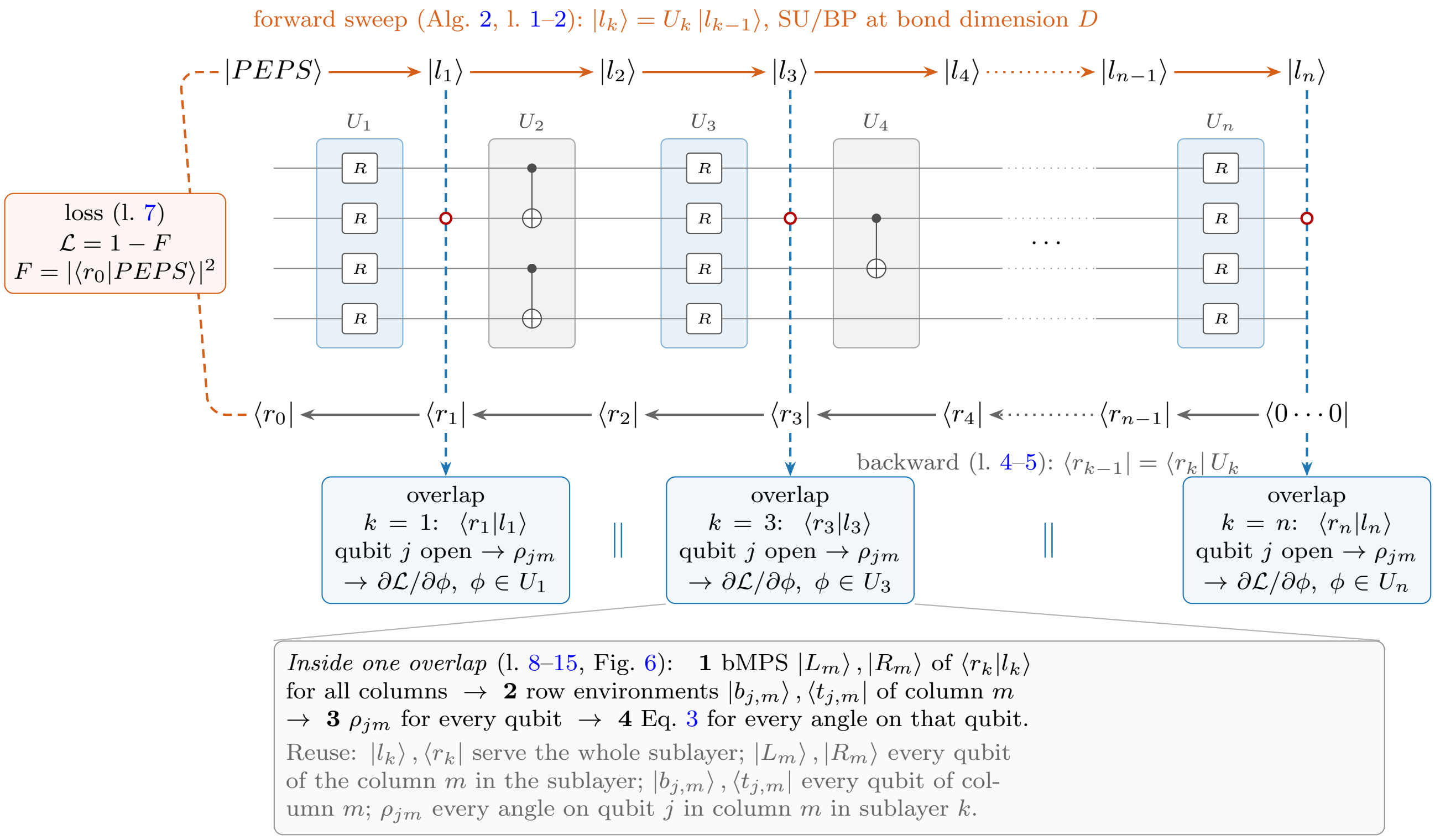
\begin{figure*}[t]
\centering
\begin{tikzpicture}[x=1.03cm,y=1cm,
  st/.style={font=\small,fill=white,inner sep=1.5pt},
  fw/.style={-{Stealth[length=5pt]},thick,acc2},
  bw/.style={-{Stealth[length=5pt]},thick,black!60},
  tap/.style={densely dashed,thick,acc},
  tapbox/.style={draw=acc,fill=acc!6,rounded corners=3pt,font=\footnotesize,align=center,inner sep=3pt,text width=2.75cm},
  rg/.style={draw=black!65,fill=white,rounded corners=1pt,font=\tiny,minimum width=0.42cm,minimum height=0.36cm,inner sep=0pt},
  lab/.style={font=\footnotesize}]
\def\wy{{0.9,0.3,-0.3,-0.9}}
\foreach \x/\t in {1/1,5/1,11/1}{\fill[acc!10,draw=acc!55,rounded corners=2pt] (\x-0.5,-1.25) rectangle (\x+0.5,1.25);}
\foreach \x in {3,7}{\fill[black!5,draw=black!35,rounded corners=2pt] (\x-0.5,-1.25) rectangle (\x+0.5,1.25);}
\foreach \x/\l in {1/{U_1},3/{U_2},5/{U_3},7/{U_4},11/{U_n}} \node[font=\scriptsize,text=black!70] at (\x,1.45) {$\l$};
\foreach \y in {0.9,0.3,-0.3,-0.9}{
  \draw[black!45] (0,\y)--(8.45,\y); \draw[black!45,dotted] (8.45,\y)--(9.55,\y); \draw[black!45] (9.55,\y)--(12,\y);}
\node[font=\small] at (9,0) {$\cdots$};
\foreach \x in {1,5,11}{\foreach \y in {0.9,0.3,-0.3,-0.9}{\node[rg] at (\x,\y) {$R$};}}
\newcommand{\cx}[3]{\fill[black!70] (#1,#2) circle (1.6pt); \draw[black!70] (#1,#2)--(#1,#3);
  \draw[black!70,fill=white] (#1,#3) circle (3.2pt); \draw[black!70] (#1-0.112,#3)--(#1+0.112,#3) (#1,#3-0.112)--(#1,#3+0.112);}
\cx{3}{0.9}{0.3}\cx{3}{-0.3}{-0.9}\cx{7}{0.3}{-0.3}
\foreach \x/\l in {0/{\ket{PEPS}},2/{\ket{l_1}},4/{\ket{l_2}},6/{\ket{l_3}},8/{\ket{l_4}},10/{\ket{l_{n-1}}},12/{\ket{l_n}}}
   \node[st] (t\x) at (\x,2.05) {$\l$};
\foreach \a/\b in {0/2,2/4,4/6,6/8,10/12} \draw[fw] (t\a)--(t\b);
\draw[fw,dotted] (t8)--(t10);
\node[lab,text=acc2,anchor=south west] at (-0.35,2.4) {forward sweep (Alg.~\ref{alg:grad}, l.~\ref{ln:fw0}--\ref{ln:fw}):\ $\ket{l_k}=U_k\ket{l_{k-1}}$, SU/BP at bond dimension $D$};
\foreach \x/\l in {0/{\bra{r_0}},2/{\bra{r_1}},4/{\bra{r_2}},6/{\bra{r_3}},8/{\bra{r_4}},10/{\bra{r_{n-1}}},12/{\bra{0\cdots0}}}
   \node[st] (b\x) at (\x,-2.05) {$\l$};
\foreach \a/\b in {2/0,4/2,6/4,8/6,12/10} \draw[bw] (b\a)--(b\b);
\draw[bw,dotted] (b10)--(b8);
\node[lab,text=black!65,anchor=north] at (9.0,-2.38) {backward (l.~\ref{ln:bw0}--\ref{ln:bw}):\ $\bra{r_{k-1}}=\bra{r_k}U_k$};
\draw[densely dashed,thick,acc2,rounded corners=6pt] (t0.west)--++(-0.45,0)--($(b0.west)+(-0.45,0)$)--(b0.west);
\node[draw=acc2,fill=acc2!6,rounded corners=3pt,font=\footnotesize,align=center,inner sep=3pt,anchor=east] at (-0.55,0)
   {loss (l.~\ref{ln:loss})\\[1pt]$\mathcal L=1-F$\\$F=|\langle r_0|PEPS\rangle|^2$};
\foreach \x/\k in {2/1,6/3,12/n}{
  \draw[tap] (t\x.south)--(b\x.north);
  \node[draw=red!70!black,fill=white,circle,inner sep=1.3pt,thick] at (\x,0.3) {};
  \node[tapbox] (o\x) at (\x,-3.55) {overlap $k=\k$: \;$\langle r_\k|l_\k\rangle$\\qubit $j$ open $\to\rho_{jm}$\\[1pt]$\to\partial\mathcal L/\partial\phi,\ \phi\in U_\k$};
  \draw[tap,-{Stealth[length=5pt]}] (b\x.south)--(o\x.north);}
\node[font=\large,text=acc] at (4,-3.55) {$\parallel$};
\node[font=\large,text=acc] at (9.0,-3.55) {$\parallel$};
\coordinate (zl) at (0.0,-4.75); \coordinate (zr) at (12.85,-4.75);
\draw[black!30] (o6.south west)--(zl) (o6.south east)--(zr);
\node[draw=black!40,fill=black!2,rounded corners=3pt,anchor=north west,inner sep=4pt,text width=13.0cm,font=\footnotesize] (zoom) at (zl)
 {\textit{Inside one overlap} (l.~\ref{ln:tap}--\ref{ln:grad}, Fig.~\ref{fig:bmps_contraction}):\quad
  \textbf{1}~bMPS $\ket{L_m},\ket{R_m}$ of $\langle r_k|l_k\rangle$ for all columns
  \;$\to$\; \textbf{2}~row environments $\ket{b_{j,m}},\bra{t_{j,m}}$ of column $m$
  \;$\to$\; \textbf{3}~$\rho_{jm}$ for every qubit
  \;$\to$\; \textbf{4}~Eq.~\ref{Eq:gradient} for every angle on that qubit.\\[2pt]
  {\color{black!60}Reuse: $\ket{l_k},\bra{r_k}$ serve the whole sublayer; $\ket{L_m},\ket{R_m}$ every qubit of the column $m$ in the sublayer; $\ket{b_{j,m}},\bra{t_{j,m}}$ every qubit of column $m$; $\rho_{jm}$ every angle on qubit $j$ in column $m$ in sublayer $k$.}};
\end{tikzpicture}
\caption{Computing the gradient (Algorithm~\ref{alg:grad}) using two PEPS sweeps through the disentangling circuit. Blue sublayers contain parametrised 1-qubit gates $R$ (products of $R_Y$ and $R_Z$), grey sublayers only the CX gates. The forward state $\ket{l_k}$ and the backward state $\bra{r_k}$ meet at the cut directly after each parametrised sublayer; contracting them with the physical index of the one qubit $j$ left open (red circle) gives $\rho_{jm}$ and, via Eq.~\ref{Eq:gradient}, the gradients of the angles in that sublayer. The overlaps at different $k$ are independent ($\parallel$) and run in parallel; the loss is computed from the overlap at the left boundary.}
\label{fig:sweep}
\end{figure*}

The complete procedure is described in detail in Algorithm~\ref{alg:grad} and Fig.~\ref{fig:sweep}. We first divide the circuit into sublayers $U=U_n\cdots U_1$ of non-overlapping 1-qubit or 2-qubit gates. A forward sweep applies the sublayers to the target, $\ket{l_k}=U_k\ket{l_{k-1}}$ with $\ket{l_0}=\ket{PEPS}$, and a backward sweep builds $\bra{r_{k-1}}=\bra{r_k}U_k$ starting from $\bra{r_n}=\bra{0\cdots0}$ (lines~\ref{ln:fw0}--\ref{ln:bw}). With Simple Update or Belief Propagation both sweeps cost $O(D^{z+1})$ per gate, where $z$ is the lattice connectivity. The infidelity is then evaluated as $1-|\langle r_0|PEPS\rangle|^2$ (line~\ref{ln:loss}). This is more accurate than $1-|\langle 0 |l_n\rangle|^2$: for shallow brick-wall circuits with $\Nbw$ layers $\bra{r_0}$ is nearly exact once $D\geq4^{\Nbw}$, 
whereas $\ket{l_n}$ generally carries larger truncation errors.

All gradients with respect to the angles of a parametrized sublayer $U_k$ follow from a single overlap network $\langle r_k|l_k\rangle$, the ``overlap'' in Fig.~\ref{fig:sweep}. We contract it with bMPS from the left and from the right (line~\ref{ln:bmps}), cut each sandwich $\bra{L_m}C_m\ket{R_m}$ into rows to obtain bottom and top environments (line~\ref{ln:env}), and combine them into $\rho_{jm}$ for every qubit $j$ of column $m$ (line~\ref{ln:rho}); Eq.~\ref{Eq:gradient} then gives the gradients of all angles of the gate acting on that qubit in sublayer $k$ (line~\ref{ln:grad}). Sublayers containing only CX gates need no contraction. Two properties make this efficient. First, every intermediate object is reused: the PEPS states by the whole sublayer, the boundary MPS by every column of the sublayer, and $\rho_{jm}$ by every angle of gate acting on qubit $j$ in sublayer $k$. Second, the PEPS overlaps of different sublayers are independent and can be computed in parallel.

\FloatBarrier
\section{Numerical experiments}
\label{sec:numQPU}
Here we numerically validate and study the performance of our improved PEPS-based quantum state preparation algorithm. Towards this end we use two test cases: the ground state of the 2D Heisenberg model on square and heavy-hex lattices, and the state generated by dynamics under the Kicked Ising model~\cite{Kim2023} on the heavy-hex lattice. The latter is defined as follows:
\begin{equation}
    U_{Kicked Ising}(\theta_{h}) = \prod_{bonds} RZZ(-\pi/2) \prod_{sites}RX(\theta_{h}),
\end{equation}
Both the 2D Heisenberg~\cite{Bruognolo2021, Jordan2008, Jiang2008}, and the Kicked Ising model at small $\theta_{h}$~\cite{Patra2024, Tindall2024} (we used $\theta_{h}=\pi/8$) can be accurately simulated with PEPS methods, providing reference values to compute fidelity and to compare the expectation values of observables, such as magnetization, staggered magnetization or energy.

\begin{figure}[!th]
\includegraphics[width= 0.95\linewidth]{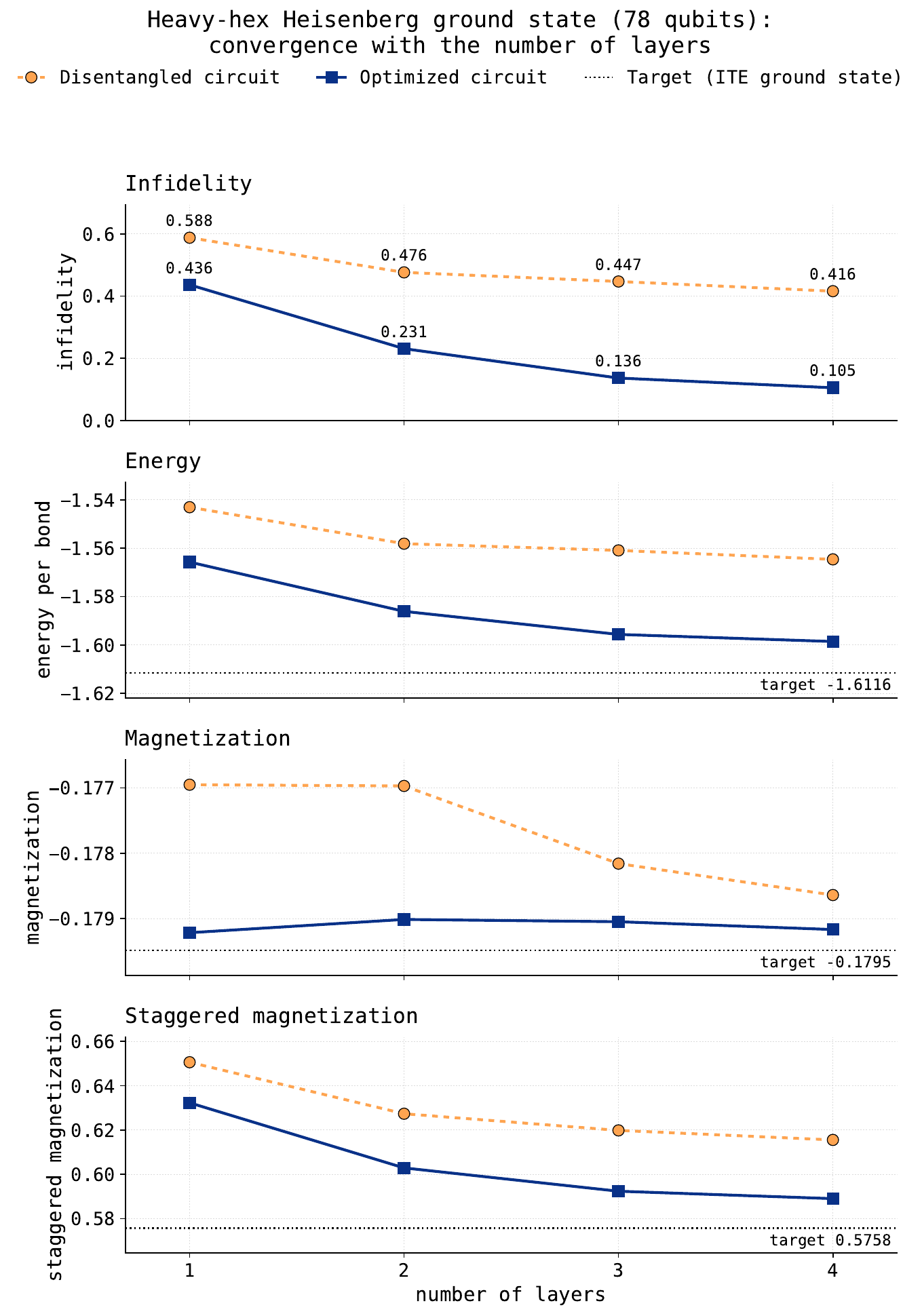} 
 \caption{\label{fig:infidelity_heavy_hex}%
   The infidelity loss, energy per bond, magnetization per site and staggered magnetization in the groundstate of the 2D Heisenberg model on the heavy-hex lattice computed in states prepared with the local disentangled circuits (orange), and the gradient-optimised ones (blue), for different number of circuit layers. The gradient optimised circuits are significantly closer to the reference values (dashed line).}
\end{figure}

\begin{figure}[!th]
\includegraphics[width= 0.95\linewidth]{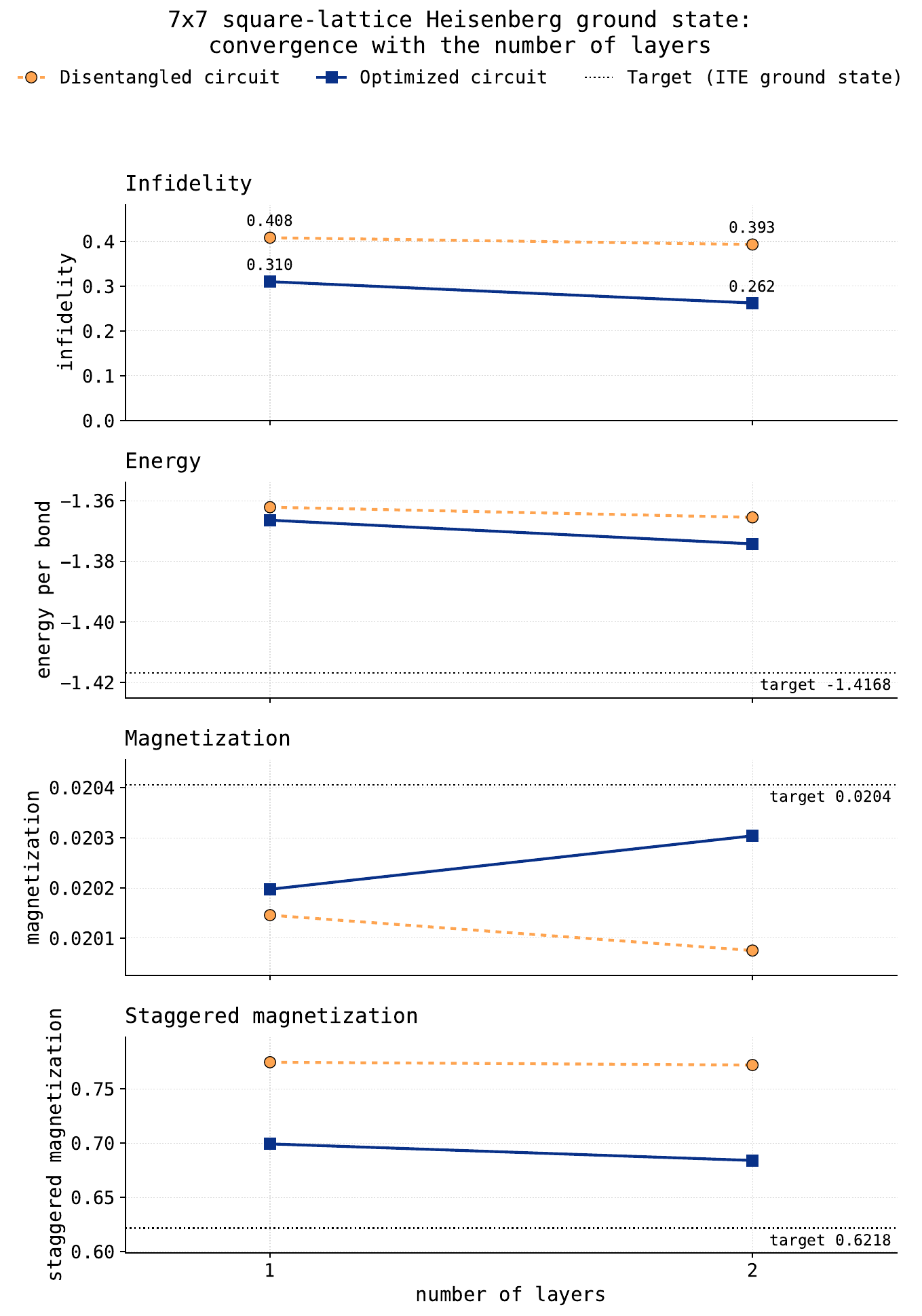} 
 \caption{\label{fig:infidelity_square}%
   The infidelity loss, energy per bond, magnetization per site and staggered magnetization in the groundstate of the 2D Heisenberg model on the $7\times7$ square lattice computed in states prepared with the local disentangled circuits (orange), and the gradient-optimised ones (blue), for different number of circuit layers. The gradient optimised circuits are significantly closer to the reference values (dashed line).}
\end{figure}

We first apply the algorithm to prepare the ground state of the 2D Heisenberg model on heavy-hex and square lattice, using a $78$ qubit patch in the topology of the IBM Heron device, and a $7\times7$ lattice patch for square case. The input PEPS states to be prepared are obtained with Simple Update PEPS evolution in imaginary time using PEPS bond dimensions $D=4$ for the heavy-hex lattice, and $D=6$ for the square lattice.



\begin{figure*}
\includegraphics[width= 0.8\linewidth]{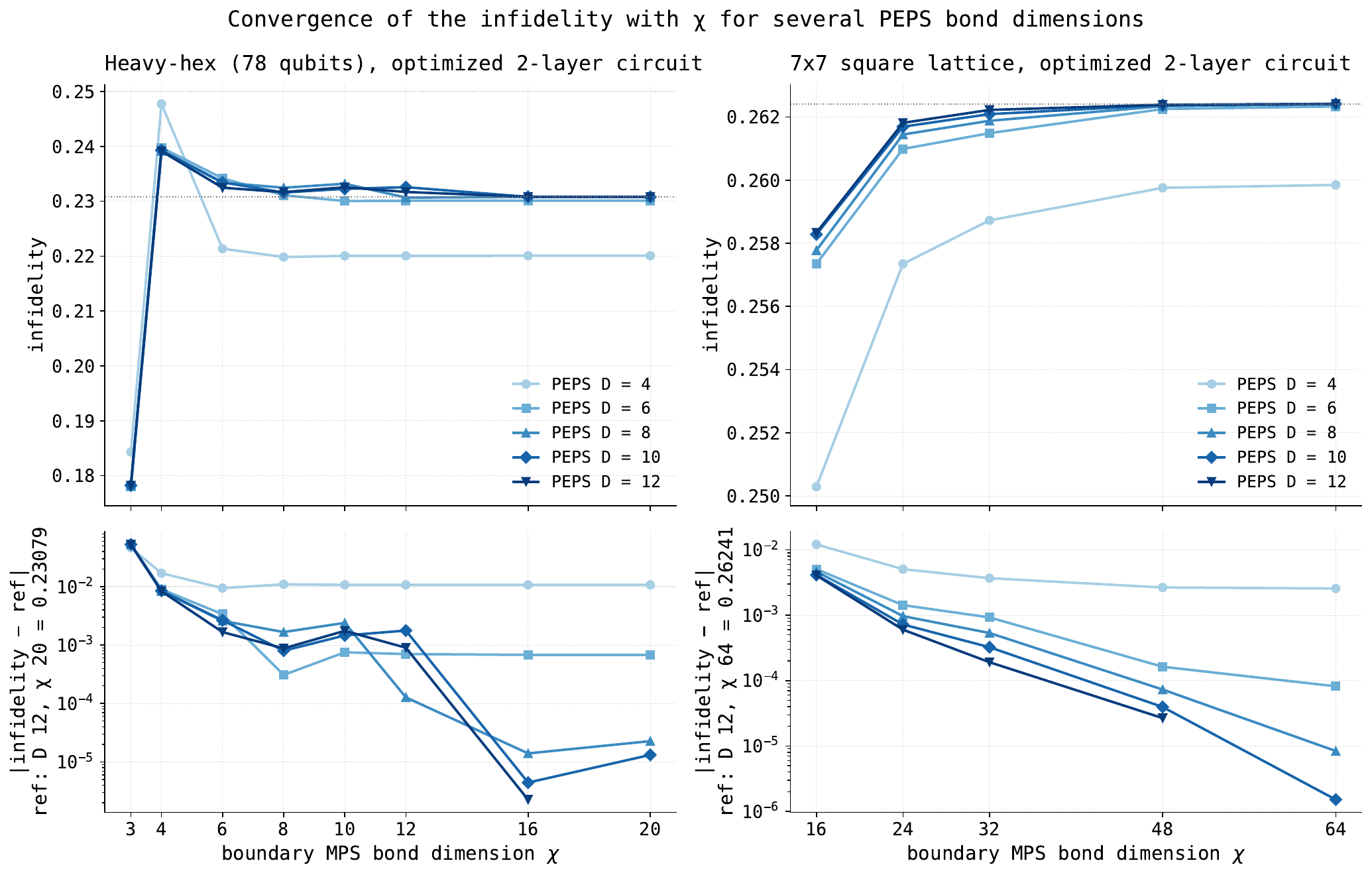} 
 \caption{\label{fig:global_loss_bond_dim}%
   Convergence of the infidelity loss with respect to the tensor network parameters - the PEPS bond dimension $D$ and the bMPS bond dimension $\chi$ for the square and heavy-hex lattices. Here we used circuits of two brick-wall layers. The results are after gradient optimisation.}
\end{figure*}

In Fig.~\ref{fig:infidelity_heavy_hex}(a) we plot the infidelity to the input state as a function of the circuit depth (the number of brick-wall layers) for the heavy-hex case. The infidelity of the globally optimized circuit obtained using our algorithm is significantly reduced: at four layers it equals $0.105$, while for the one obtained  with the local disentenglement procedure it is $0.416$. Moreover, this fidelity gap only increases with the number of circuit layers.

In Figs.~\ref{fig:infidelity_heavy_hex}(b-d) we further plot the expectation value of energy, as well as magnetization and staggered magnetization. In all of them the gradient optimized circuits are substantially closer to the reference values than the local disentanglement ones, and improve with additional circuit layers. The same quantities for the $7\times7$ square lattice are shown in Fig.~\ref{fig:infidelity_square} and are also consistently closer to the reference values for the gradient-optimized circuits. In this case we only show the results with one and two brick-wall layers, as accurate optimization for three brick-wall layers required large bond dimensions. 

In the above, the optimization on the heavy-hex lattice used bond dimensions $D=8, \chi=14$, and the square lattice results were trained with $D=8, \chi=24$ for 1 brick-wall layer and $D=12, \chi=36$ for 2 brick-wall layers. All of the results are fully converged with respect to PEPS and bMPS parameters $(D, \chi)$: we increased these until the changes in the loss and observables fall below $10^{-4}$.

We next study the convergence of the infidelity loss, and of the gradients themselves, as a function of PEPS bond dimension $D$, the boundary MPS bond dimensions $\chi$, and the number of gradient iterations. The convergence of infidelity relative to the reference state in both bond-dimension parameters is shown in Fig.~\ref{fig:global_loss_bond_dim} for both the heavy-hex and square cases. Here we used a circuit of two brick-wall layers.

\begin{figure*}
\includegraphics[width= 0.8\linewidth]{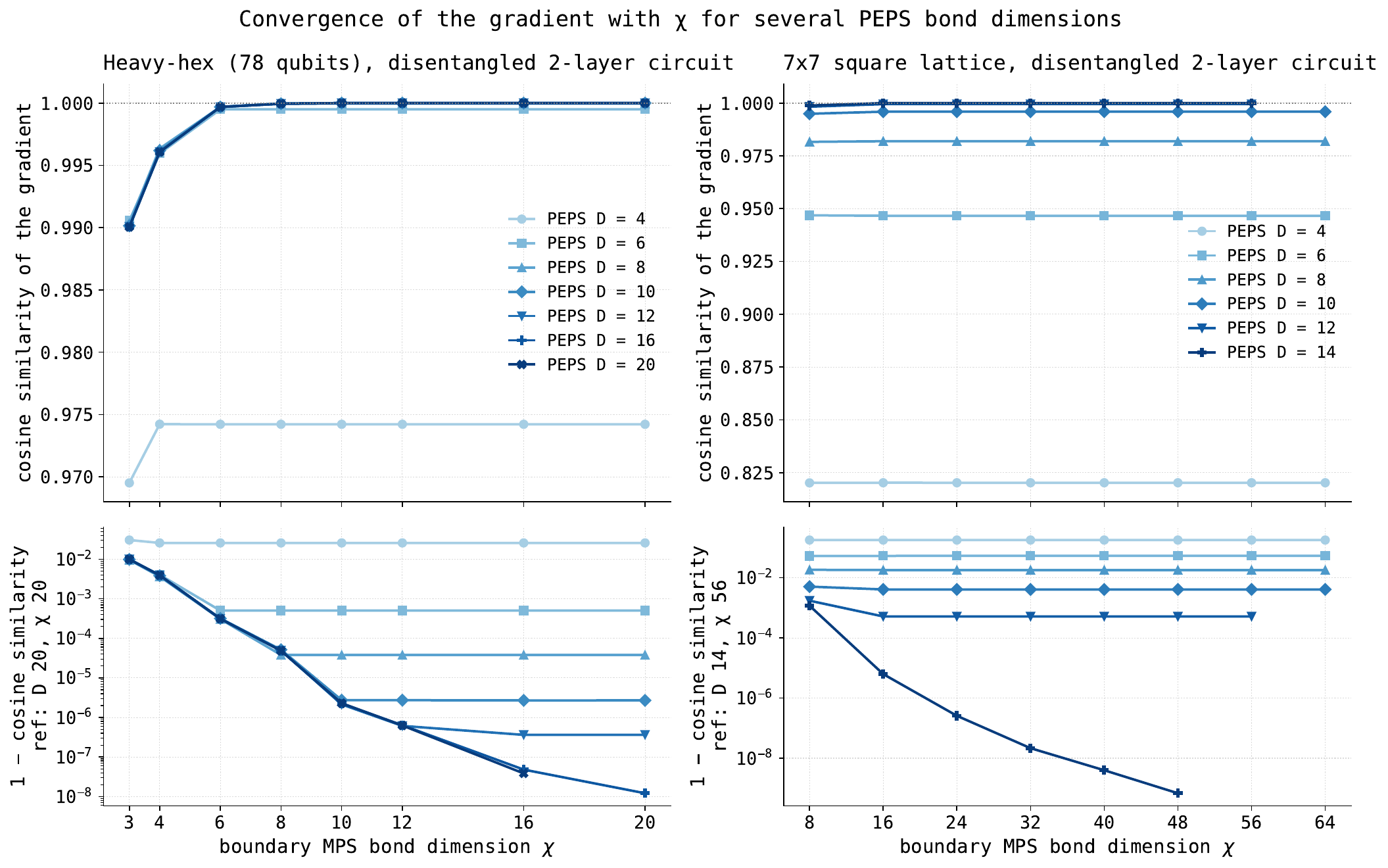} 
 \caption{\label{fig:grad_bond_dim}%
   Convergence of the gradients of the infidelity loss with respect to the tensor network parameters - the PEPS bond dimension $D$ and the bMPS bond dimension $\chi$. Convergence is measured as a cosine of the angle between the gradient at given bond dimensions and the most accurate available point (cosine similarity). The results are presented for $2$ layer brick-wall circuits after disentanglement but before gradient optimization, when gradients are still large.}
\end{figure*}

In Fig.~\ref{fig:grad_bond_dim} we plot the convergence of the gradients as a function of the bond dimensions, also for a two-layer circuit, for both the heavy-hex and square lattices. We used the cosine similarity metric measuring distance to the most accurate available value, and measured the gradients at the beginning of the optimization procedure, when they are still large. All of the plots show a rapid convergence of the gradient estimates with the PEPS and bMPS bond dimensions. Additionally, in Fig.~\ref{fig:grad_check} we show that the gradient estimates converge when measured with respect to exact state vector simulations, which we validated on $20$ qubits on the $4 \times 5$ square lattice, and on a $21$-qubit $2$-loop subgraph of the heavy-hex lattice.



 An interesting difference between the square and heavy-hex lattices we observed is that on the latter comparably small bMPS bond dimensions $\chi$ are required for both loss and gradients' convergence: $\chi=12-14$ was sufficient. This is in stark contrast to the usual empirical rule suggesting $\chi \approx D^{2}$. On the square lattice, on the other hand, we usually required larger $\chi \approx 30$ for successful optimization in our computations. 
 Moreover, the bMPS contraction on the square lattice has a higher computational cost as a function of $D$. Together, these make circuit optimization on the square lattice much more difficult compared to the heavy-hex case. 
 

\begin{figure}
\includegraphics[width= 0.9\linewidth]{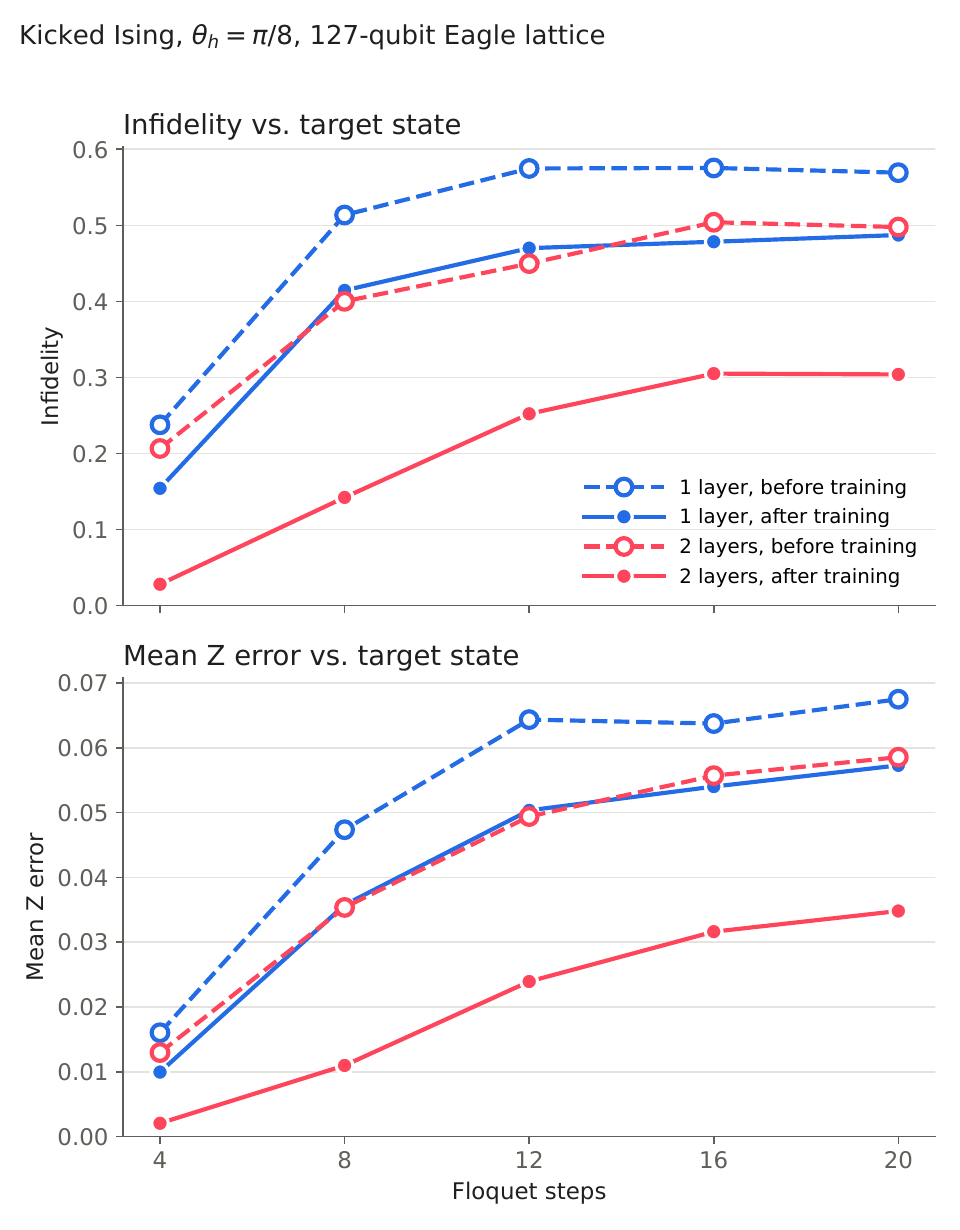} 
 \caption{\label{fig:kicked_ising}%
   State preparation of the 127-qubit Floquet-evolved Kicked Ising model on the heavy-hex lattice. We plot the infidelity and the error in the average $\langle Z \rangle$ expectation values of states at different Floquet step using the locally disentangled, or globally gradient-tuned circuits. In blue we show results for 1-layer preparation circuits, in red for two layers. In both cases the gradient-tuned circuits are substantially more accurate. As expected, for a fixed number of circuit layers the accuracy decreases as the time-evolved states becomes more complex.}
\end{figure}

Finally, we consider the example of the kicked Ising model, on the 127-qubit heavy-hex topology of the Eagle family of IBM devices used in Ref.~\cite{Kim2023} (which can be trivially embedded into that of IBM Heron).
Here we used $D=10$ PEPS states to simulate dynamics for $4, 8, 12, 16$ and $20$ Floquet layers, providing input states from which the quantum circuits were synthesized. For gradient-based optimization we used $D=10$ and $\chi=16$. In all cases except $4$ Floquet steps and $2$ brick-wall layers, the trained brick-wall circuit has a lower depth than the original evolution circuit (which was simulated with PEPS). In Fig.~\ref{fig:kicked_ising}, we plot the infidelity loss of the locally disentangled and gradient-based circuits, and the mean error in local $Z$ observables between the target state and the one state produced by the trained circuit, for one and two brick-wall layers. As in the Heisenberg ground state case, our gradient-trained circuits are consistently more accurate in both state fidelity and expectation values. Unsurprisingly, the accuracy increases with the number of layers, but for a fixed number of layers it decreases with the number of Floquet steps, as the quantum state becomes more complex.

\section{Conclusions and Outlook}
\label{sec:conclusions}

We developed a novel method of PEPS-based quantum state preparation and gradient computation of variational circuits on a PEPS simulator. We use local disentanglement of PEPS to return a quantum circuit, which serves as a warm-start to a global, gradient-based optimization step, which is made possible by our new efficient tensor-network gradient computation procedure. 

We validated the performance and convergence of the method preparing the ground states of the 2D Heisenberg model on the heavy-hex and square topologies corresponding to the existing superconducting QPUs, and also preparing states obtained from a simulation of the Kicked Ising Floquet dynamics. In simulations our globally optimised circuits systematically achieve significantly higher fidelity and accuracy in observable estimation at the same circuit depth than those obtained with local disentangling circuits.
We further observed a significant difference in computational cost of preparation between the square and heavy hex-lattice, with the latter being much easier to optimise. This is in line with the results of Ref.~\cite{Rudolph2025}, which points to the particular difficulty which smaller-loop lattices pose for tensor network methods. 




There are several open research directions remaining. We used the examples of magnetic ground state, which are close to criticality, but still have local entanglement. It would be interesting to use disentanglement and gradient optimization to prepare long-range entangled topological ordered states, such as $RVB$ liquid ~\cite{Verstraete2006}, toric code ground state or double-semion topological order ground state ~\cite{Gu2009, Buerschaper2009}. These states are not expected to be preparable with brick-wall circuits we used, but more general sequential circuits may succeed. Another potential application concerns a general method of data loading on a quantum computer, first encoding the classical data in a PEPS, and then preparing the corresponding quantum circuit as described here. The first step of this procedure is missing, however: while for MPS and tree tensor networks tensor cross interpolation (TCI)~\cite{Oseledets2010, Dolgov2020, Ritter2024, NunezFernandez2025, Tindall2024tree} allows to encode arbitrary data in the tensor, there is no TCI analogue for PEPS. A possible route may involve converting a MPS to a PEPS using the algorithm of Ref. ~\cite{Zaletel2020}.


\FloatBarrier
\bibliographystyle{unsrtnat}   
\bibliography{ref.bib}

\newpage
\appendix
\section{Technical details}

\begin{figure}
\includegraphics[width= 0.9\linewidth]{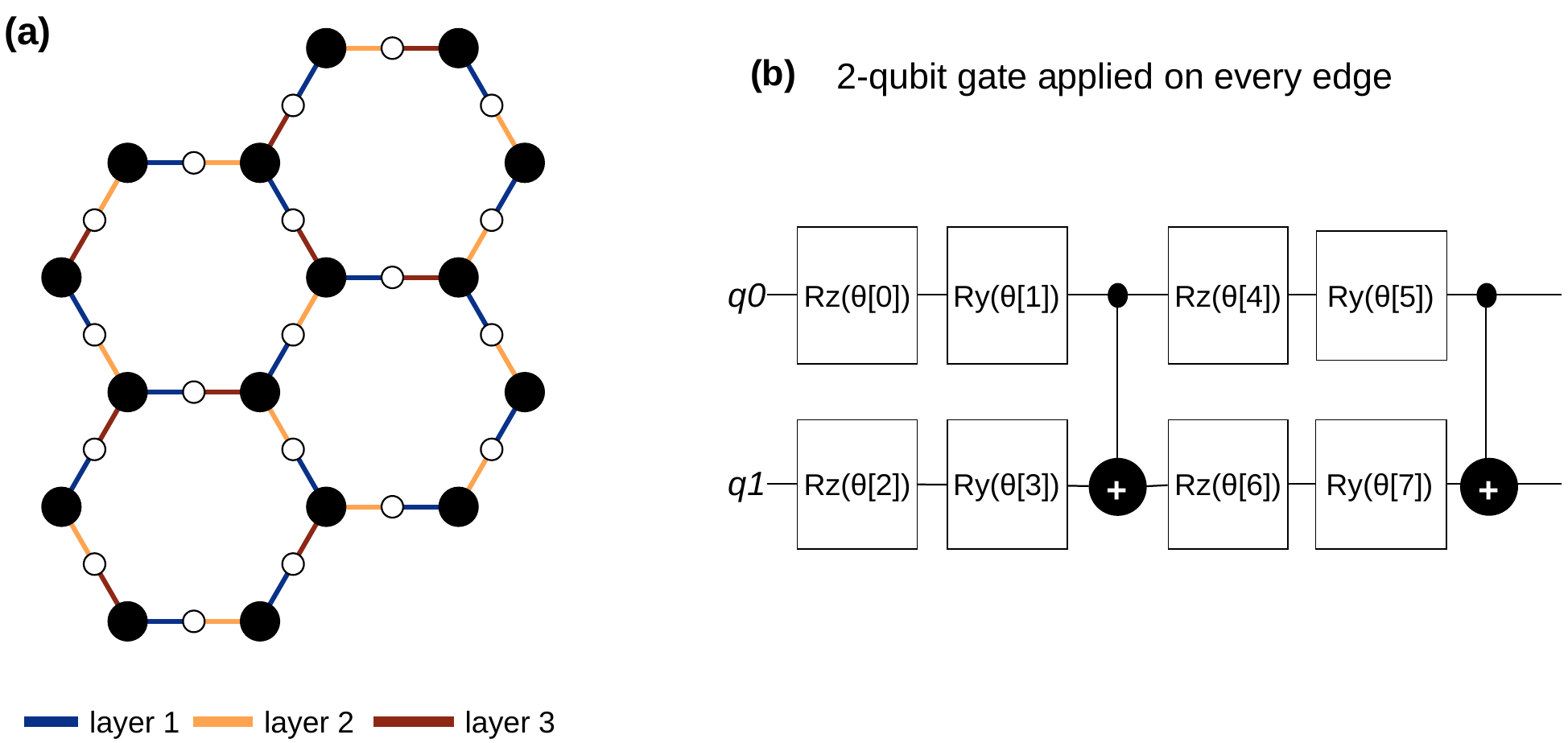} 
 \caption{\label{fig:brick-wall}%
   The structure of the circuit ansatz used for disentanglement. (a) we first color the bonds of the lattice in such a way that bonds of the same color do not share the vertices. The 2-qubit gates can be applied simultaneously to bonds of the same color. Brick-wall layer is then built by sequentially applying 2-qubit gates from different colour classes. (b)  We use the following parametrised circuit ansatz for the 2-qubit disentanglement block.   }
\end{figure}

\begin{figure}
\includegraphics[width= 1.0\linewidth]{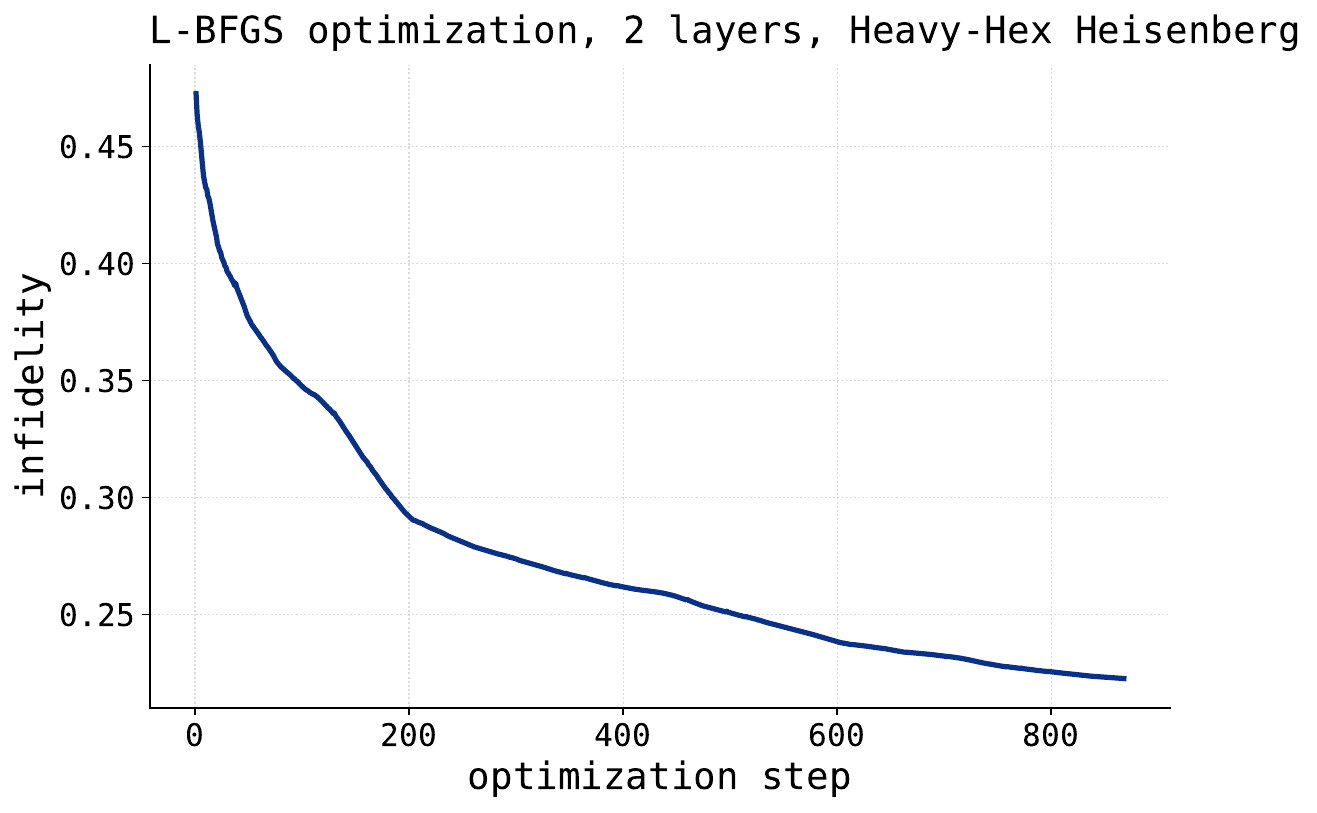} 
 \caption{\label{fig:lbfgs}%
   Convergence of the infidelity loss with the number of L-BFGS optimization steps.    }
\end{figure}

\begin{figure*}
\includegraphics[width= 0.8\linewidth]{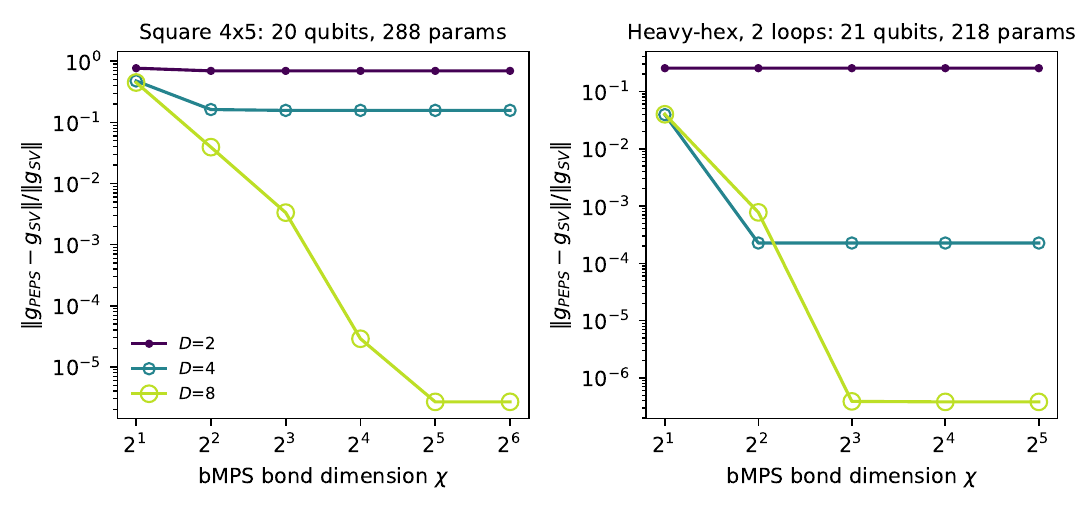} 
 \caption{\label{fig:grad_check}%
   Convergence of the infidelity loss gradients $g_{PEPS}$ with respect to the tensor network parameters - the PEPS bond dimension $D$ and the bMPS bond dimension $\chi$. Here we compare to the exact state vector based gradient $g_{SV}$ computation at small system sizes. For the square lattice we used  the ground state of the 2D Heisenberg model on the $4\times 5$ lattice as the target state, while for the heavy-hex case we used a small $21$ qubit subgraph (consisting of two loops) of the heavy-hex lattice.      }
\end{figure*}

Here we provide some of the technical details of the numerical procedures and experiments performed.

In Fig.~\ref{fig:brick-wall}(a) we show the example of lattice edge colouring, which defines pairs of qubits, on which the gates can be applied simultaneously. This is a single sublayer of the ansatz circuit, the complete set of which form a complete layers. The circuit we considered in our numerical experiments were up to four layers deep. In Fig.~\ref{fig:brick-wall}(b) we show a parametrised circuit ansatz for a single block acting on a pair of qubits in a sublayer. It consists of CX gates, and parametrised single qubit rotations.

In addition to the gradient convergence results in Fig.~\ref{fig:grad_bond_dim} we also show convergence with respect to exact state vector simulations in small system. In Fig.~\ref{fig:grad_check} we do this for the Heisenberg model on $20$ qubits on the $4 \times 5$ square lattice, and on a $21$-qubit $2$-loop subgraph of the heavy-hex lattice.

We also investigated the convergence of the algorithm as a function of the number of optimizer steps. In Fig. ~\ref{fig:lbfgs} we plot is for the heavy-hex Heisenberg model optimization. Initially, the loss quickly decreases and most of the loss change can be attributed to the first $200$ loss evaluations. After this, the loss change is much slower. As the calculation of PEPS gradients is slow, in most cases it is therefore better to stop optimization after a fixed number of steps. Note that if the bond dimensions $(D, \chi)$ are insufficient for accurate gradient computation, these become effectively noisy. In such a case, optimizers such as ADAM may be more efficient than L-BFGS we used.

\end{document}